\documentclass[%
 reprint,
 superscriptaddress,
nofootinbib,
 amsmath,amssymb,
 aps,
 prd,
]{revtex4-2}

\usepackage{graphicx}                   
\usepackage{dcolumn}                    
\usepackage{bm}                         
\usepackage{hyperref}                   
\usepackage{physics}                    
\usepackage[noabbrev]{cleveref}         
\usepackage{braket}                     
\usepackage{mathtools}                  
\usepackage{siunitx}                    
\usepackage[caption=false]{subfig}      

\newcommand*{\SU}[1][2]{\ensuremath{\text{SU}\qty(#1)}}
\DeclareMathOperator{\arcosh}{arccosh}

\begin{document}

\preprint{}

\title{Vector boson scattering from the lattice}

\author{Patrick Jenny}
 \email{patrick.jenny@edu.uni-graz.at}
\author{Axel Maas}
 \email{axel.maas@uni-graz.at}
\author{Bernd Riederer}
 \email{bernd.riederer@uni-graz.at}
\affiliation{Institute of Physics, NAWI Graz, University of Graz, Universitätsplatz 5, A-8010 Graz, Austria}

\date{\today}

\begin{abstract}
We study vector-boson scattering of the physical, gauge-invariant states in a reduced standard-model setup on the lattice for various parameter sets. To this end, the phase shift in the scalar channel is determined using a L\"uscher-type analysis. The results can be readily interpreted in terms of the Higgs properties and a reunitarized Fr\"ohlich-Morchio-Strocchi analysis at Born level. The only deviation appears for a Higgs mass below the elastic threshold, where we find a negative scattering length indicative of the bound-state nature of the physical scalar degree of freedom. We assess the possible implications for an experimental detection of the effect.
\end{abstract}

\maketitle

\section{\label{sec:introduction}Introduction}

Electroweak physics is very successfully described in the framework of the Brout-Englert-Higgs (BEH) effect using perturbation theory \cite{ParticleDataGroup:2020ssz,Bohm:2001yx}. However, from a formal point \cite{Banks:1979fi,Frohlich:1980gj,Frohlich:1981yi,Osterwalder:1977pc} of view the physical degrees of freedom cannot be the elementary states charged under the weak interactions; they change under weak gauge transformations, are thus gauge-dependent and hence unphysical. Rather, composite operators and thus effectively bound states should be the correct asymptotic degrees of freedom \cite{Banks:1979fi,Frohlich:1980gj,Frohlich:1981yi}. This seeming paradox is resolved by the Fr\"ohlich-Morchio-Strocchi (FMS) mechanism \cite{Frohlich:1980gj,Frohlich:1981yi}.

The FMS mechanism is essentially an expansion of the composite operators in terms of the Higgs vacuum expectation value (vev). It shows that to leading order the properties of the bound states are identical to the ones of the elementary states \cite{Frohlich:1980gj,Frohlich:1981yi}. As a result of this, especially the masses and widths should agree. This has been supported by several lattice studies \cite{Maas:2012tj,Maas:2013aia,Maas:2020kda,Shrock:1986fg}. Similar statements also hold for leptons \cite{Frohlich:1980gj,Frohlich:1981yi,Afferrante:2020fhd,Lee:1988ut}, and even hadrons \cite{Egger:2017tkd,Maas:2017wzi,Fernbach:2020tpa}. For an overview and additional background on this issue we refer to the review \cite{Maas:2017wzi}.

But the off-shell properties and interactions of the bound states can in principle deviate at higher orders of the vev \cite{Maas:2012tj,Egger:2017tkd,Maas:2020kda,Dudal:2020uwb,Dudal:2021dec}. Correspondingly, such deviations have been observed for the form factor of the physical vector bosons in lattice simulations \cite{Maas:2018ska} and analytically for off-shell Higgs \cite{Maas:2020kda,Dudal:2020uwb} and vector boson properties \cite{Dudal:2021dec}. This can have potentially observable consequences.

We explore here the impact on one of the most central processes in electroweak physics, namely vector-boson scattering (VBS) \cite{Lee:1977eg,Denner:1996ug,Denner:1997kq,Buarque:2021dji,Covarelli:2021gyz}, using lattice simulations. However, for various reasons \cite{Maas:2017wzi} the full standard model cannot yet be fully simulated on the lattice and therefore we consider a reduced standard model to this end. Furthermore we explore different values of the masses and coupling constants to study the behavior in a more general way. The theoretical setup of this work is detailed in \cref{sec:ewh-sector}. Further details of the lattice simulations are described in \cref{sec:lattice}, and can be skipped if only the comparison to analytical calculations are of interest.  

The ultimate result of our lattice calculations is the phase shift in the spin zero partial wave in the elastic region, shown in section \ref{sec:results}. Additionally we obtain the fully gauge-invariant prediction for this quantity at leading order using the FMS mechanism in \cref{sec:pt}. Finally we compare both results in \cref{sec:results}. Provided the Higgs mass is below the elastic threshold we have a parameter-free analytical calculation. It is found that it agrees well with the data, except close to the threshold. Near the threshold, we observe a negative scattering length, which is a sign of a bound state of finite extent. The observed scale is consistent with the one inferred from the form factor of the vector bosons themselves \cite{Maas:2018ska}, drawing a nicely consistent picture. We also estimate the consequences at the level of a cross-section, and discuss how this could erroneously fake a composite Higgs signal from beyond the standard model. Finally, if there is no Higgs below threshold, we find that we can use the results to identify the presence and properties of a Higgs resonance.

We summarize the results in \cref{sec:wrapup}, in which we draw a conclusive picture of VBS in a fully gauge-invariant setting. Additional technical details are relegated to appendices \ref{ap:spectrum} to \ref{ap:pt}.

\section{\label{sec:ewh-sector}Theoretical setup}

\subsection{A reduced standard model}

We study a reduced standard model setup \cite{Montvay:1994cy} which consists of the weak interaction \SU[2]$_W$ gauge theory coupled to a complex scalar doublet $\phi$. The corresponding Lagrangian of this model is given by
\begin{equation}\label{eqn:l_ew}
    \mathcal{L} = -\frac{1}{4}W_{\mu\nu}^{a}W^{a\,\mu\nu} + \qty(D_{\mu}\phi)^{\dagger}\qty(D^{\mu}\phi) - \lambda\qty(\phi^{\dagger}\phi - f^2)^2
\end{equation}
with $W_{\mu}$ the gauge fields, $W_{\mu\nu}^{a}$ the usual field-strength tensor and $D_{\mu}$ the usual covariant derivative, where the latter two depend on the gauge coupling $g$.

For latter convenience we also introduce a matrix representation of the scalar field $\phi$ by
\begin{equation}\label{eqn:scalar_mat}
    \Phi=\mqty(\phi_1 & -\phi_2^{\dagger} \\ \phi_2 & \phi_1^{\dagger})
\end{equation}
where $\phi_i$ are the corresponding components of the complex doublet. In this form gauge transformations act on $\Phi$ as a left multiplication. Additionally there is a global \SU[2] symmetry, which we will call the custodial symmetry in the following and therefore denote as \SU[2]$_\mathcal{C}$. This symmetry acts on $\Phi$ as a right multiplication.

The potential term in \cref{eqn:l_ew} allows for a non-trivial minimum and thus for a BEH effect to take place. The usual procedure \cite{Bohm:2001yx,Maas:2017wzi} is then to select a particular minimum (``\emph{spontaneous gauge-symmetry breaking}'') by gauge-fixing, e.g.\ to a 't Hooft gauge, and then perform a shift $\phi\to v+\eta$, where $|v|=f$ is the Higgs vev. This leads to three degenerate massive gauge bosons $W$ of mass $m_{W}=gf/2$ and one massive scalar of mass $m_{H} = \sqrt{2\lambda f^2}$. The massive scalar, associated to the component of $\eta$ along $v$, will be called the Higgs, while the remaining three degrees of freedom are orthogonal to $v$ and act as would-be Goldstone bosons. The degeneracy of the gauge bosons is enforced by the diagonal subgroup\footnote{In a purely perturbative setting this diagonal subgroup is often called the custodial group. However, since its definition is dependent on the choice of gauge, we reserve this name here for the physically observable global group \SU[2]$_\mathcal{C}$.} of \SU[2]$_W\times$\SU[2]$_\mathcal{C}$.

The parameters of \cref{eqn:l_ew} can then be fixed arbitrarily. In our present work we are interested in either SM-like settings, i.e.\ a Higgs somewhat heavier than the $W$, or an even heavier Higgs to understand how resonances can be described. The latter is also of relevance for 2-Higgs doublet models \cite{Branco:2011iw}, in which very similar considerations hold as in the SM \cite{Maas:2016qpu}. We therefore fix the $W$ mass throughout to \SI{80.375}{\GeV}, and vary the Higgs mass. We also choose somewhat larger weak gauge couplings than in the SM, which parametrically amplify the results, and thus create a better signal-to-noise ratio in the lattice simulations.

\subsection{\label{subsec:states}Elementary fields vs.\ physical particles}

In a perturbative setup the above described degrees of freedom are then used as the asymptotic states \cite{Bohm:2001yx}. However, because the shift and the BEH effect required gauge-fixing, which indeed is the only possibility to do so, these are gauge-dependent states, and thus formally unphysical \cite{Frohlich:1980gj,Frohlich:1981yi,Maas:2012ct}. The usual Becchi-Rouet-Stora-Tyutin (BRST) construction is actually insufficient to identify the physical degrees of freedom, which is a consequence of the combination of the existence of Gribov copies at arbitrary weak coupling \cite{Fujikawa:1982ss} and Haag's theorem \cite{Haag:1992hx}, see \cite{Maas:2017wzi} for a review.

To overcome this issue, it is therefore necessary to use intrinsically gauge-invariant objects (e.g.\ composite operators of the elementary fields like $\phi^{\dagger}\phi$) as the physical degrees of freedom \cite{Banks:1979fi,Frohlich:1980gj,Frohlich:1981yi}. In general these objects need to be treated nonperturbatively, but due to the presence of the BEH-effect, it is possible to augment perturbation theory in a consistent way. This is the so-called FMS mechanism \cite{Frohlich:1980gj,Frohlich:1981yi,Maas:2017wzi}.

In practice, this is a two-step process. Given any manifestly gauge-invariant composite operator, first insert the usual BEH split in a convenient gauge. In this way, a (connected) correlation function like a propagator becomes a sum of (individually gauge-dependent) correlation functions, e.g.\
\begin{gather}
    \langle (\phi^{\dagger}\phi)^\dagger(x)\phi^{\dagger}\phi(y)\rangle=\langle(v\eta)^\dagger(x)v\eta(y)\rangle\nonumber\\
    +\langle(\eta^\dagger\eta)^\dagger(x)v\eta(y)+x\leftrightarrow y\rangle+\langle(\eta^\dagger\eta)^\dagger(x)(\eta^\dagger\eta)(y)\rangle.
\end{gather}
In a second step a double expansion in $v$ and the other coupling constants can be made. E.g.\ at leading order in $v$, the propagator of the composite operator $\phi^{\dagger}\phi$ therefore coincides to all orders in all other couplings with the elementary Higgs propagator, and especially has the same mass and width.

Especially, in the SM the leading order in $v$ recovers the ordinary perturbative results\footnote{In general, this is not the case and there can be a qualitative discrepancy, already at the level of the spectrum, depending on the specific interplay of the gauge group and the custodial group \cite{Maas:2017wzi, Maas:2014pba, Torek:2018qet, Sondenheimer:2019idq}.} \cite{Maas:2017wzi}.

But this also implies that higher orders in $v$ will lead to deviations from the results of ordinary perturbation theory. Indeed, such deviations have previously been observed on \cite{Maas:2018ska} and off \cite{Egger:2017tkd,Maas:2020kda,Fernbach:2020tpa,Reiner:2021bol,Dudal:2020uwb} the lattice. In addition, because the FMS mechanism is still perturbative in nature, it is not able to describe genuine non-perturbative surplus bound states, e.g.\ internal excitations. Both contributions can potentially be misidentified in experiments as new physics, if not thoroughly taken into account \cite{Maas:2017wzi,Egger:2017tkd,Fernbach:2020tpa}.

Recent studies showed that higher orders in $v$ can be accounted for analytically \cite{Maas:2017wzi,Maas:2020kda,Dudal:2020uwb,Dudal:2021dec}. Testing the quality of this, as well as ensuring that no relevant surplus effects beyond this double-expansion arise, requires non-perturbative methods. Both are aims of the present work.

\subsection{\label{subsec:operators}Particles and the scattering process}

As discussed above, we need composite operators which describe the real physical observables, and thus the particles observed in experiments. To make contact to the successful perturbative treatment we also need their connection to the elementary fields.

In our reduced SM all particles can be classified by their $J^{P}$ quantum numbers, with spin $J$ and parity $P$. The gauge-invariant operators carry an additional quantum number, describing the transformation properties under the global symmetry. Therefore these will be denoted by $J^{P}_{\mathcal{C}}$, with $\mathcal{C}$ being the custodial representation. The elementary fields likewise carry a representation of the gauge group and, in case of the scalar, also of the custodial group.

In general it is possible to construct operators with arbitrary $J^{P}_{\mathcal{C}}$, see e.g.\ \cite{Wurtz:2013ova,Sondenheimer:2019idq,Dobson:2021sgl}. Here we aim at investigating the elastic region in the scalar singlet channel. Therefore only $J^{P}_0$ states and the states of the single open decay channel are needed, and thus the lightest stable particle into which a $0^{+}_0$ could decay. That has been established to be the $1^{-}_{1}$ state \cite{Wurtz:2013ova,Maas:2014pba}. Thus, the process we are looking at is elastic scattering of two custodial triplet vector bosons, the non-perturbatively and manifestly gauge-invariant version of the elementary VBS process \cite{Denner:1996ug,Denner:1997kq,Buarque:2021dji,Covarelli:2021gyz}. In this way \cite{Maas:2017wzi}, Haag's theorem \cite{Haag:1992hx} is manifestly preserved and the weak-coupling consequences of the Gribov-Singer ambiguity \cite{Fujikawa:1982ss,Singer:1978dk} are evaded. 

This process is particularly interesting as it is considered to be a sensitive probe to new physics \cite{Buarque:2021dji,Covarelli:2021gyz,Gallinaro:2020cte}, and deviations due to unaccounted-for SM background would therefore have potentially a substantial impact on searches for new physics.

Gauge-invariant composite operators with suitable overlap with the ground states, and thus the physical versions of the asymptotic ``Higgs''- and ``W/Z''-states, are given by \cite{Frohlich:1980gj,Frohlich:1981yi,Maas:2013aia,Wurtz:2013ova,Maas:2014pba}
\begin{align}\label{eqn:physical_higgs}
    \mathcal{O}^{0^{+}_{0}} &= \phi^{\dagger}\phi = \frac{1}{2}\tr\qty{\Phi^{\dagger}\Phi} \\
    \label{eqn:physical_vector}
    \mathcal{O}^{1^{-}_{1}\, a}_{\mu} &= \tr\qty{\tau^{a}\Phi^{\dagger}D_{\mu}\Phi}
\end{align}
with $\Phi$ as defined in \cref{eqn:scalar_mat} and $\tau^a=\sigma^a/2$ the \SU[2]-generators proportional to the Pauli matrices $\sigma^a$. In the FMS expansion to NLO these states are found to expand to the elementary Higgs and gauge bosons, and thus have the same mass \cite{Maas:2020kda,Dudal:2020uwb}. This is confirmed by lattice results \cite{Maas:2013aia,Maas:2020kda}.

We therefore have now the necessary setup to  consider the physical VBS process. As non-perturbative approach we will use lattice methods to determine the phase shift in the $J^P=0^+$ partial wave in the elastic region, i.e.\ for $2m_W\le \min(4m_W,2m_H)$. To this end, we will use a L\"uscher analysis \cite{Gattringer:2010zz,Luscher:1990ck} detailed in \cref{sec:lattice}. The results of the L\"uscher analysis do not rely on any interpretation in terms of the FMS approach, and are thus a stand-alone outcome of our work. In particular, we determine the scattering lengths to obtain an estimate of the bound state scale for the lattice settings where we do have a stable scalar. Otherwise, we will test for the presence of resonances in the elastic region.

Thereafter, we will determine the same phase shift entirely analytically at reunitarized Born level in the FMS expansion, as detailed in section \ref{sec:pt}. This is again a result in itself giving now a fully perturbative description of the physical VBS process. Finally, we will compare this analytical result to the lattice result, and can therefore test the validity of the FMS expansion and the possible consequence of deviations as our final result.

\section{\label{sec:lattice}Lattice methods}

\subsection{Configurations}

The configurations for this work have been created using the methods described in \cite{Maas:2013aia,Maas:2014pba,Riederer:2020tvj} implementing the unimproved lattice Wilson action corresponding to \cref{eqn:l_ew} \cite{Montvay:1994cy}. In the following $U_{\mu}(x)$ will be used to denote the lattice gauge links, while the same symbols $\phi\qty(x)$, and $\Phi\qty(x)$ are used for the lattice versions of the scalar fields.

\begin{table}[t!]
    \centering
    \caption{Parameter sets. The lattice spacing has been set by fixing the mass of the lightest state in the $1^-_1$ channel to \SI{80.375}{\GeV} \cite{Maas:2013aia,Riederer:2020tvj}, which induces the lowest possible error, being at the sub-percent level. The running weak gauge coupling in the MiniMOM scheme \cite{vonSmekal:2009ae} has been determined as in \cite{Maas:2013aia} with an error of order \SI{1}{\percent}.}
    \label{tab:parameters}
    \begin{tabular}{c|c|c|c|c|c|c}
        Name & $\beta$ & $\kappa$ & $\gamma$ & $\alpha_{W,\SI{200}{\GeV}}$ & $a^{-1}\ \qty[\si{\GeV}]$ & $m_H\ \qty[\si{\GeV}]$ \\\hline\hline
        Set 1 &	2.7984	&	0.2984	&	1.317	&	0.492	&	287 & $148^{+6}_{-20}$ \\
        Set 2 & 2.8859	&	0.2981	&	1.334	&	0.448	&	291 & $149^{+6}_{-11}$\\
        Set 3 & 4.0000	&	0.2850	&	0.970	&	0.219	&	289 & -- \\
        Set 4 & 4.0000	&	0.3000	&	1.000	&	0.211	&	243 & $275^{+3}_{-3}$
    \end{tabular}
\end{table}

The lattice parameters are given in \cref{tab:parameters}. We note that the theory appears to have only very mild finite-volume and discretization effects \cite{Maas:2014pba} due to the absence of an exceptionally light state like the pion in QCD. But the theory has also much fewer resonances or bound states than QCD \cite{Wurtz:2013ova,Maas:2014pba}. To have enough states within the elastic region for a determination of the phase shifts therefore requires that at least some scattering states must lie within this window. In addition, the bosonic states yield a lot of statistical noise\footnote{We explicitly opted not to use over-smearing \cite{Wurtz:2013ova} to avoid loosing information of potentially excited states.}, requiring very large statistics (\cref{tab:statistics}), and thus limiting the physical volume. Thus, we were forced to use relatively coarse lattices, see \cref{tab:parameters} and lattice sizes $8^4$, $12^4$, $16^4$, $20^4$, $24^4$, $28^4$, and $32^4$. This setup allowed to reach the goals of this work.

We note that our lattice sets use larger weak gauge couplings than we would have in a SM setting. This and the Higgs mass are therefore our main control parameter. As mentioned, this choice of weak gauge coupling appears to amplify the expected effects, and therefore allows us to extract them from the noise. As far as has been systematically investigated \cite{Maas:2014pba,Maas:2018ska,Afferrante:2020fhd} non-trivial effects seem to not depend qualitatively on the weak gauge coupling, but of course quantitatively. A logical extension of this work in the future, using substantially more computing time, is therefore to move towards weaker gauge coupling to assess the quantitative size of the effects better.

\subsection{\label{subsec:spectroscopy}Spectroscopy}

The ingredients we need for the determination of the phase shifts are the infinite-volume masses of the involved particles as well as the energy levels as a function of the volume within the elastic window \cite{Briceno:2017max,Luscher:1990ux}.

To obtain them, we use a variational analysis approach \cite{Gattringer:2010zz}. To optimize overlap, we employ a very large operator basis. To this end, we construct 36 primary operators in the relevant $0^+_0$ channel, which we then smear up to four times using an APE smearing as detailed in \cite{Maas:2014pba,Philipsen:1996af}, generating in total 180 operators, see \cref{eqn:op_basis1,eqn:op_basis2}.

We start with the basic operators
    \begin{gather}
        \mathcal{O}_{H}\qty(x) = \phi^{\dagger}\qty(x)\phi\qty(x) \label{eqn:higgsop}\\
        \mathcal{O}_{W}\qty(x) = \Tr\qty{U_{\mu}\qty(x)U_{\nu}\qty(x+e_{\mu})U_{\mu}^{\dagger}\qty(x+e_{\nu})U_{\nu}^{\dagger}\qty(x)} \label{eqn:wballop}\\
        \mathcal{O}_{0^{+}_{n}}\qty(x) = \sum_{\mu=1}^{3}\Tr\qty{\frac{\Phi^{\dagger}\qty(x)}{\sqrt{\det(\Phi\qty(x))}}U_{\mu}\qty(x)\frac{\Phi\qty(x+e_{\mu})}{\sqrt{\det(\Phi\qty(x+e_{\mu}))}}} \label{eqn:0plusopnorm}\\
        \mathcal{O}_{0^{+}}\qty(x) = \sum_{\mu=1}^{3}\Tr\qty{\Phi^{\dagger}\qty(x)U_{\mu}\qty(x)\Phi\qty(x+e_{\mu})} \label{eqn:0plusop}\\
        \mathcal{O}_{1^{-}_{n}\mu}^{a}\qty(x) = \Tr\qty{\tau^{a}\frac{\Phi^{\dagger}\qty(x)}{\sqrt{\det(\Phi\qty(x))}}U_{\mu}\qty(x)\frac{\Phi\qty(x+e_{\mu})}{\sqrt{\det(\Phi\qty(x+e_{\mu}))}}} \label{eqn:1minusopnorm}\\
        \mathcal{O}_{1^{-}\mu}^{a}\qty(x) = \Tr\qty{\tau^{a}\Phi^{\dagger}\qty(x)U_{\mu}\qty(x)\Phi\qty(x+e_{\mu})} \label{eqn:1minusop}
    \end{gather}
    with \crefrange{eqn:higgsop}{eqn:0plusop} in the $0^{+}_{0}$ channel and \crefrange{eqn:1minusopnorm}{eqn:1minusop} in the $1^{-}_{1}$ channel. They can be interpreted as the following physical objects
    \begin{itemize}
        \item $\mathcal{O}_{H}$ describes a two-Higgs bound-state considered as the physical Higgs particle like \cref{eqn:physical_higgs}
        \item $\mathcal{O}_{W}$ describes a W-ball and expands to $1-W_{\mu\nu}^a W^{a\mu\nu}$ in the continuum limit
        \item $\mathcal{O}_{1^{-}\mu}^{a}$ is the vector triplet and thus is considered as the physical W-Boson as in \cref{eqn:physical_vector}
        \item $\mathcal{O}_{0^{+}_{\qty(n)}}$ and $\mathcal{O}_{1^{-}_{\qty(n)}\mu}^{a}$ have no direct physical connections but yield a very stable signal \cite{Evertz:1985fc}
    \end{itemize}
    These interpretations have to be understood in the sense of a large overlap between the lattice operators and the physical states \cite{Maas:2014pba}. One needs to keep in mind, that potentially all operators in a channel contribute to all states.

    From the local operators the momentum space $\mathcal{O}\qty(\va{p})$ versions have then been obtained by a lattice Fourier transformation. From these operators we constructed the following operator basis used for the variational analysis in the $1^{-}_{1}$ channel
    \begin{equation}\label{eqn:op_basis_vec}
        \mathcal{O}^{1^{-}_{1}\,a}_{1-10\mu}  = \left\{
        \begin{gathered}
            \mathcal{O}_{1^{-}\mu}^{\qty(0-4)a}\qty(\va{0}) \\
            \mathcal{O}_{1^{-}_{n}\mu}^{\qty(0-4)a}\qty(\va{0})
        \end{gathered}\right.
    \end{equation}
    where the $\qty(0-4)$ refers to the different smearing levels.
    The basis used is rather small, but based on \cite{Maas:2014pba} sufficient to reliably determine the ground state mass. For the $0^{+}_{0}$ channel the basis is chosen much larger:
    \begin{equation}\label{eqn:op_basis1}
        \mathcal{O}_{1-90}^{0^{+}_{0}}  = \left\{
        \begin{aligned}
            &\left.
            \begin{gathered}
                \mathcal{O}_{W}^{\qty(0-4)}\qty(\va{p}) \\
                \mathcal{O}_{H}^{\qty(0-4)}\qty(\va{p}) \\
                \mathcal{O}_{0^{+}}^{\qty(0-4)}\qty(\va{p}) \\
                \mathcal{O}_{0^{+}_{n}}^{\qty(0-4)}\qty(\va{p}) \\
                \mathcal{O}_{1^{-}\mu}^{\qty(0-4)a}\qty(-\va{p}) \, \mathcal{O}_{1^{-}\mu}^{\qty(0-4)a}\qty(\va{p}) \\
                \mathcal{O}_{1^{-}_{n}\mu}^{\qty(0-4)a}\qty(-\va{p}) \, \mathcal{O}_{1^{-}_{n}\mu}^{\qty(0-4)a}\qty(\va{p}) \\
            \end{gathered}\right\} \text{such that }\qty|\va{p}|^2=0
            \\
            &\left.
            \begin{gathered}
                \mathcal{O}_{W}^{\qty(0-4)}\qty(-\va{p}) \, \mathcal{O}_{W}^{\qty(0-4)}\qty(\va{p}) \\
                \mathcal{O}_{H}^{\qty(0-4)}\qty(-\va{p}) \, \mathcal{O}_{H}^{\qty(0-4)}\qty(\va{p}) \\
                \mathcal{O}_{0^{+}}^{\qty(0-4)}\qty(-\va{p}) \, \mathcal{O}_{0^{+}}^{\qty(0-4)}\qty(\va{p}) \\
                \mathcal{O}_{0^{+}_{n}}^{\qty(0-4)}\qty(-\va{p}) \, \mathcal{O}_{0^{+}_{n}}^{\qty(0-4)}\qty(\va{p}) \\
                \mathcal{O}_{1^{-}\mu}^{\qty(0-4)a}\qty(-\va{p}) \, \mathcal{O}_{1^{-}\mu}^{\qty(0-4)a}\qty(\va{p}) \\
                \mathcal{O}_{1^{-}_{n}\mu}^{\qty(0-4)a}\qty(-\va{p}) \, \mathcal{O}_{1^{-}_{n}\mu}^{\qty(0-4)a}\qty(\va{p}) \\
            \end{gathered}\right\} \text{such that }\qty|\va{p}|^2=1,2
        \end{aligned}\right.
    \end{equation}
     The different absolute momentum values need to be understood in ascending order, i.e. $\mathcal{O}_{31-60}^{0^{+}_{0}}$ corresponds to $\qty|\va{p}|^2=1$. In addition to the operators above interpolators containing two (or more) particle have also been considered
    \begin{equation}\label{eqn:op_basis2}
        \mathcal{O}_{91-180}^{0^{+}_{0}} = \qty(\mathcal{O}_{1-90}^{0^{+}_{0}})^2
    \end{equation}
    to find the inelastic threshold.

We construct then the full correlation matrix as
    \begin{gather}\label{eqn:correlation_matrix}
        C_{ij}\qty(\Delta t) = \frac{1}{L_t} \sum_{t=0}^{L_t-1}C_{ij}\qty(t,\Delta t)\nonumber\\
        C_{ij}\qty(t,\Delta t) = \expval{\left[\mathcal{O}_{i}\qty(t)-\expval{\mathcal{O}_{i}\qty(t)}\right]\left[\mathcal{O}_{j}\qty(t + \Delta t)-\expval{\mathcal{O}_{j}\qty(t+\Delta t)}\right]}
    \end{gather}
    with $L_t$ the temporal extent and the standard errors given by
    \begin{equation}\label{eqn:error_correlation_matrix}
        \Delta C_{ij}\qty(\Delta t) = \sqrt{\frac{1}{L_t\qty(L_t-1)}\sum_{t=0}^{L_t-1}\qty[C_{ij}\qty(t,\Delta t) - C_{ij}\qty(\Delta t)]^2}
    \end{equation}
in both channels. However, including all 180 operators in the $0^{+}_{0}$ channel, which are partly extremely noisy, did not yield a stable numerical result. We therefore did a preselection and postselection of the operators included, the latter only for the $0^+_0$ channel. In the following we sketch the most important steps of the analysis performed, skipping most technicalities. More details are given in \cref{ap:spectrum}.

First, we noted that including smaller smearing levels did not show any statistically significant indications of overlap with further states, but did increase the statistical errors. We therefore reduced the set to the 4-times smeared operators, leaving 36 and two operators in the $0^+_0$ and $1^-_1$ channels, respectively. In the $1^-_1$ channel, this was sufficient to obtain the necessary stable ground state signal.

In the $0^+_0$ channel we considered a variable number of operators $n$ for every setting and volume. We continuously increased our operator basis from one operator to all 36 operators. These where added according to their time-summed relative errors, starting with the least noisiest. This ensured to keep the overall noise at a sustainable level. We then monitored the resulting spectrum as a function of the operator number, also in comparison to the expected non-interacting spectrum. We limited the number of operators to the point where the spectrum was stable under addition of further operators and the levels showed a volume-development which was consistent with the physical one, i.e.\ except for possible avoided level crossings a polynomial or exponential behavior. In this way we identified 10-20 operators for every volume and lattice parameter set, which we admitted to the final analysis

To determine the actual energy levels, we performed an arbitrary precision\footnote{In this way no preconditioning of the correlator matrix was necessary or would have had any effect.} variational analysis on the correlator matrix. We identified levels by sorting the eigenvalues in terms of the maximum overlap of the eigenvectors at finite time with those at time zero.

However, the result was still very noisy, and most eigenvalues did not have a signal for the whole time extent of the lattice. Fits using single or double $\cosh$ behavior of the effective masses did not yield satisfactory results. On the other hand, the usual search for plateaus in the effective energy\footnote{Note that in standard literature this is usually referred to as the effective mass but in the present context it is more accurate to use the term energy.} of a correlator defined as \cite{Gattringer:2010zz}
\begin{equation}\label{eqn:effective_energy}
    E_{eff}(t) = -  \frac{1}{t -\frac{L_t}{2}}\arcosh\qty( \frac{C(t)}{C(L_t/2)})
\end{equation}
was also not possible, as the necessary quantity $C(L_t/2)$ was usually drowned in noise.

To circumvent the problem, we used a predictor for $C(L_t/2)$ \cite{Jenny:2021}. Assuming that the variational analysis yielded eigenvalues which became dominated by a single state within a few time slices, we determined a value of $C(L_t/2)$. Therefore we used the assumption that $E(t)$ plateaus for every single level, and then averaged over a window of time slices where this assumption turned out to be reasonable. We then used the resulting averaged prediction of $C(L_t/2)$ to determine in a constant fit the final value of the mass. We used extensive tests with mock-up data at various error levels to validate this method and optimize our window, which in the end was $[2,L_t/2-2]$ for sufficiently large volumes \cite{Jenny:2021}. The resulting energy levels were remarkably stable across different sets of operators and volumes, giving further confidence in the method.

We determined errors for our energy levels by varying the correlator matrix as $C_{ij}\to C_{ij}\pm\Delta C_{ij}$ and repeating the full analysis three times, yielding asymmetric error bars. Changing the amount of statistics showed that this was a suitable estimate of the actual statistical errors.

In the $1^-_1$ channel the ground state mass obtained by this method were extremely precise, with errors at the per mill level. We fitted the ground state with an exponential fall-off to extract the infinite volume mass. The results are listed in \cref{ap:parameters}. When there was a signal for a bound state  below the elastic threshold in the $0^+_0$ channel, we did so likewise, though here the errors are much larger. The final result will be shown in \cref{subsec:spectra}, and the ground state mass is given in \cref{tab:parameters} for the two cases in which it was below the threshold. In addition \cref{tab:parameters} also contains the resonance mass obtained for data set 4 as will be explained in \cref{subsec:phase shift}.

\subsection{\label{subsec:luescher}Phase shifts and the L\"uscher analysis}

In the determination of the phase shifts \cite{Briceno:2017max,Luscher:1990ux} we follow closely the steps in \cite{Lang:2011mn,Lang:2012db}. The L\"uscher analysis is used to map the energy levels determined in section \ref{subsec:spectroscopy} to the phase shift in the partial waves of the corresponding quantum numbers, here $0^+_0$. We note in passing that we have a symmetric situation as the two-particle states in the elastic region are made up of identical particles.

To this end, the energy in the center of mass frame for states in the  elastic region on the lattice are given by
\begin{align}\label{eqn:lattice_energy}
    \cosh{\qty(\frac{E}{2})} &= \cosh{\qty(m)} + 2\sin{\qty(\frac{\abs{\va{p}}}{2})}^2
\end{align}
in lattice units and $m$ the infinite volume mass, replacing the usual continuum dispersion relation $E =2\sqrt{m^2+\va{p}^2}$. We used the lattice dispersion relation throughout but note, as in \cite{Lang:2011mn,Lang:2012db}, that the difference compared to using the continuum one was almost irrelevant.

Interaction distorts the levels and therefore a generalized momentum is introduced which describes the deviation from the lattice momenta without interaction,
\begin{equation}
    q =\abs{\va{p}} \frac{L}{2\pi}.
\end{equation}
These momenta are related to the phase shift $\delta_J$ by a transcendental equation of the type $\tan{\qty(\delta_J\qty(q))}=f\qty(q)$ \cite{Luscher:1990ux}. For our case of $J=0$ and vanishing center-of-mass momentum the defining equation can be expressed as \cite{Luscher:1990ux}
\begin{equation}\label{eqn:phase shift}
    \tan(\delta_{0}\qty(q)) = \frac{\pi^{\frac{3}{2}}q}{\mathcal{Z}_{00}^{\va{0}}\qty(1,q^2)}
\end{equation}
with
\begin{gather}\label{eqn:zeta}
    \mathcal{Z}_{Jm}^{\va{d}}\qty(r,q^2) = \sum_{\va{x}\in P_{\va{d}}} \frac{\qty|\va{x}|^{J}Y_{Jm}\qty(\va{x})}{\qty(\va{x}^2-q^2)^{r}} \\\nonumber
    P_{\va{d}} = \Set{\va{x} \in \mathbb{R}^3 | \va{x} = \va{y} + \frac{\va{d}}{2}, \va{y} \in \mathbb{Z}^3 }
\end{gather}
where $Y_{Jm}$ are the usual spherical harmonics.

To make practical use of this connection an analytical continuation of the transcendental function $\mathcal{Z}_{Jm}^{\va{d}}$ is necessary. For this purpose we use the prescription detailed in appendix \ref{ap:zeta}, see also \cite{CP-PACS:2007wro,Lang:2011mn}.

The phase shift can also be used to characterize \cite{Sakurai:2011zz} the nature of the scattering. Defining near the threshold the scattering length $a_0$ by \cite{Lang:2012db}
\begin{equation}
    \abs{\va{p}}\cot\qty(\delta_{0}\qty(s)) = \frac{2\mathcal{Z}_{00}^{\va{0}}\qty(1,q^2)}{\sqrt{\pi}L} = \frac{1}{a_0} +\order{\abs{\va{p}}^2} \label{eqn:scattering_length},
\end{equation}
where a negative value indicates scattering on a bound state with a characteristic scale given by $a_0$. A positive or zero value indicates the absence of such a bound state.

We again propagate errors in the phase shift analysis by performing it three times with the input energy levels varied within $\pm 1\sigma$.

\section{\label{sec:pt}Perturbative description}

\begin{figure*}
\centering
\includegraphics[width=0.97\linewidth]{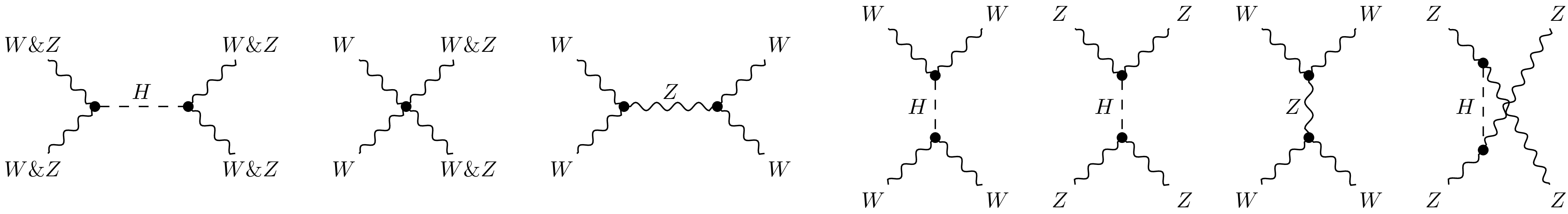}
\caption{Possible transition diagrams at tree-level. The ampersand (\&) means that all possible combinations need to be considered here. E.g.\ the second diagram contains $WW\rightarrow WW$ and $WW\rightarrow ZZ$.}
\label{fig:transitionelements}
\end{figure*}

In the usual perturbative description of VBS \cite{Lee:1977eg,Denner:1996ug,Denner:1997kq,Buarque:2021dji,Covarelli:2021gyz} the initial and final states are fully specified. In the present case, however, we do not have access to the precise initial and final states, except knowing that it is made up by two $1^-_1$ states in an $s$-wave with zero total momentum and net zero custodial charge.

To determine the relevant matrix element we therefore need to consider a mixture of all admissible states. We construct it as a tensor product using Clebsch-Gordan coefficients in spin and custodial charge. This yields the state
\begin{align}\label{eqn:instate}
    \ket{in} = \ket{0^+_0} =\frac{1}{3} \sum_{m,c = -1}^1 \qty(-1)^{m+c} \ket{1^-_1;m,c}\otimes\ket{1^-_1;-m,-c}.
\end{align}

The full transition amplitude for the elastic region is then obtained as
\begin{align}
    \mathcal{M} =& \mel{out}{T}{in} = \frac{1}{9} \sum_{\substack{m_1,m_2\\c_1,c_2}=-1}^1 \qty(-1)^{m_1+m_2+c_1+c_2} \nonumber\\\label{eqn:transition}
    &\mel{m_1,c_1;-m_1,-c_1}{T}{m_2,c_2;-m_2,-c_2}
\end{align}
which is basically a sum over all 81 possible full 4-point vertices $\ev{\mathcal{O}_{m_1,c_1}\mathcal{O}_{-m_1,-c_1}\mathcal{O}_{m_2,c_2}\mathcal{O}_{-m_2,-c_2}}$. This quantity is still fully gauge-invariant and would need again non-perturbative methods to be evaluated.

To obtain an analytical prediction, we apply now the FMS mechanism to it. Performing first an expansion of the operators in the vev to leading order yields that $\mathcal{O}_{m,c}$ are replaced by $W_{m,c}$\footnote{Switching between a $\pm$ or a $1,2$ description is only a rotation of basis, and will not alter the final matrix elements. Therefore these will be omitted and $WW$ refers to any valid combination like $W^+W^-$ or $W^1W^2$. Despite the $Z$ is in our degenerate case just $W^3$, we revert here momentarily to the notation with a $Z$ for better comparability to the standard results.}, up to an irrelevant prefactor, which can be absorbed in the definition of the operators \cite{Frohlich:1981yi,Maas:2017wzi}. At this order, the matrix elements of scattering process are thus the usual perturbative ones \cite{Egger:2017tkd,Maas:2018ska}.

Performing now the perturbative expansion finally leaves us with a sum over all possible polarised transition-amplitudes at tree-level for the processes $WW\rightarrow WW$, $WW\leftrightarrow ZZ$ and $ZZ\rightarrow ZZ$. Additionally, due to CPT- and CP-symmetry some helicity amplitudes can be related to each other \cite{Denner:1996ug,Denner:1997kq}. This reduces our problem to 12 different amplitudes, which are listed in \cref{tab:ap_amplitudes}. The full amplitude in a Born-level approximation for the transition-amplitude in \cref{eqn:transition} thus reads
\begin{widetext}
\begin{alignat}{5}
\mathcal{M}_B =\frac{1}{9} \Big[
    4& \mel{\text{WW}}{T}{\text{WW}} _{\{0,0,0,0\}}
    &&-16 \mel{\text{WW}}{T}{\text{WW}} _{\{\pm ,\mp ,0,0\}}
    &+8& \mel{\text{WW}}{T}{\text{WW}} _{\{\pm ,\mp ,\pm ,\mp \}}
    &+8& \mel{\text{WW}}{T}{\text{WW}} _{\{\pm ,\mp ,\mp ,\pm \}} \nonumber\\
    -4& \mel{\text{WW}}{T}{\text{ZZ}} _{\{0,0,0,0\}}
    &&+16 \mel{\text{WW}}{T}{\text{ZZ}} _{\{\pm ,\mp ,0,0\}}
    &-8& \mel{\text{WW}}{T}{\text{ZZ}} _{\{\pm ,\mp ,\pm ,\mp \}}
    &-8& \mel{\text{WW}}{T}{\text{ZZ}} _{\{\pm ,\mp ,\mp ,\pm \}} \nonumber\\
    +&\mel{\text{ZZ}}{T}{\text{ZZ}} _{\{0,0,0,0\}}
    &&-4 \mel{\text{ZZ}}{T}{\text{ZZ}} _{\{\pm ,\mp ,0,0\}}
    &+2& \mel{\text{ZZ}}{T}{\text{ZZ}} _{\{\pm ,\mp ,\pm ,\mp\}}
    &+2& \mel{\text{ZZ}}{T}{\text{ZZ}} _{\{\pm ,\mp ,\mp ,\pm\}}
    \Big]
    \label{eqn:matrixelement}
\end{alignat}
with $\mel{c_1c_2}{T}{c_3c_4}_{\{m1,m2,m3,m4\}}\equiv\mel{m_1,c_1;m_2,c_2}{T}{m_3,c_3;m_4,c_4}_{\text{tl.}}$ the tree-level amplitudes. Kinematical details like momentum assignments are given in appendix \ref{ap:pt}.
\end{widetext}

The individual amplitudes in \cref{eqn:matrixelement} can be found in the literature. Here we are following the same kinematic assignments as in \cite{Denner:1996ug,Denner:1997kq} where the pure $W$- and pure $Z$-transitions can be found. The remaining $WW\leftrightarrow ZZ$ processes can be deduced from the Feynman-Diagrams in \cref{fig:transitionelements} using e.g.\ \cite{Bohm:2001yx}. In addition, also all other possible tree-level contributions to the VBS-process are shown in \cref{fig:transitionelements}.

To obtain now an expression for the phase shift $\delta_0$ in the $s$-wave one needs the relation
\begin{align}\label{eqn:partial_waves}
    \mathcal{M}&=16\pi \sum_J(2J+1)f_J P_J(\cos\theta)\\\label{eqn:partial_amplitudes}
    f_J &= \frac{1}{32\pi(2J+1)}\int_{-1}^{1} \mathcal{M} P_J(\cos\theta)\dd{\cos\qty(\theta)} = \nonumber\\
    &= e^{i\delta_{J}}\sin\qty(\delta_J)
\end{align}
where $P_J$ are the Legendre polynomials. By projecting \cref{eqn:matrixelement} with $P_0$ yields therefore the phase shift at Born level.

However, because at Born level there is no branch cut in the amplitudes the optical theorem is violated such a phase shift would be complex. To avoid this requires either to go to higher order or to reunitarize the result. As higher orders in the present case otherwise change little and the result is much more transparent, we chose here reunitarization using the ``Direct $T$-matrix Unitarization'' approach \cite{Kilian:2014zja}. Therefore we first calculate the partial wave transition amplitude $f_J$ in the specific channel by projection with $P_0$ as mentioned above, followed by the corresponding reunitarization prescription \cite{Kilian:2014zja}
\begin{equation}\label{eqn:reunitarization}
    \mathcal{F}_J=\frac{1}{\Re\qty(\frac{1}{f_J})-i}.
\end{equation}
This is necessary to ensure unitarity, and is usually found to be only a minor effect at the energies relevant here \cite{Kilian:2014zja}. This yields then the reunitarized Born-level FMS prediction for the bound-state scattering process on the lattice, and it agrees to this order to the perturbative VBS scattering prediction. The latter would only change once higher orders in the vev would be taken into account.

To compare our results it also is instructive to obtain an expression for $\tan\qty(\delta_J)$. This can be done by solving \cref{eqn:partial_amplitudes} which yields
\begin{equation}\label{eqn:pert_tan}
    \tan\qty(\delta_J) = \frac{\mathcal{F}_J}{1+i\mathcal{F}_J} = \frac{1}{\Re\qty(\frac{1}{f_J})}.
\end{equation}
We note that the reunitarization procedure in \cref{eqn:reunitarization} simplifies the relation for $f_J\in\mathbb{R}$ to $ \tan\qty(\delta_J) = f_J$. This relation therefore holds for the Born level amplitude which is always real.

\section{\label{sec:results}Results}

\subsection{\label{subsec:spectra}Energy spectra}

\begin{figure*}
    \centering
    \subfloat[Set1]{\includegraphics[width=0.971\columnwidth]{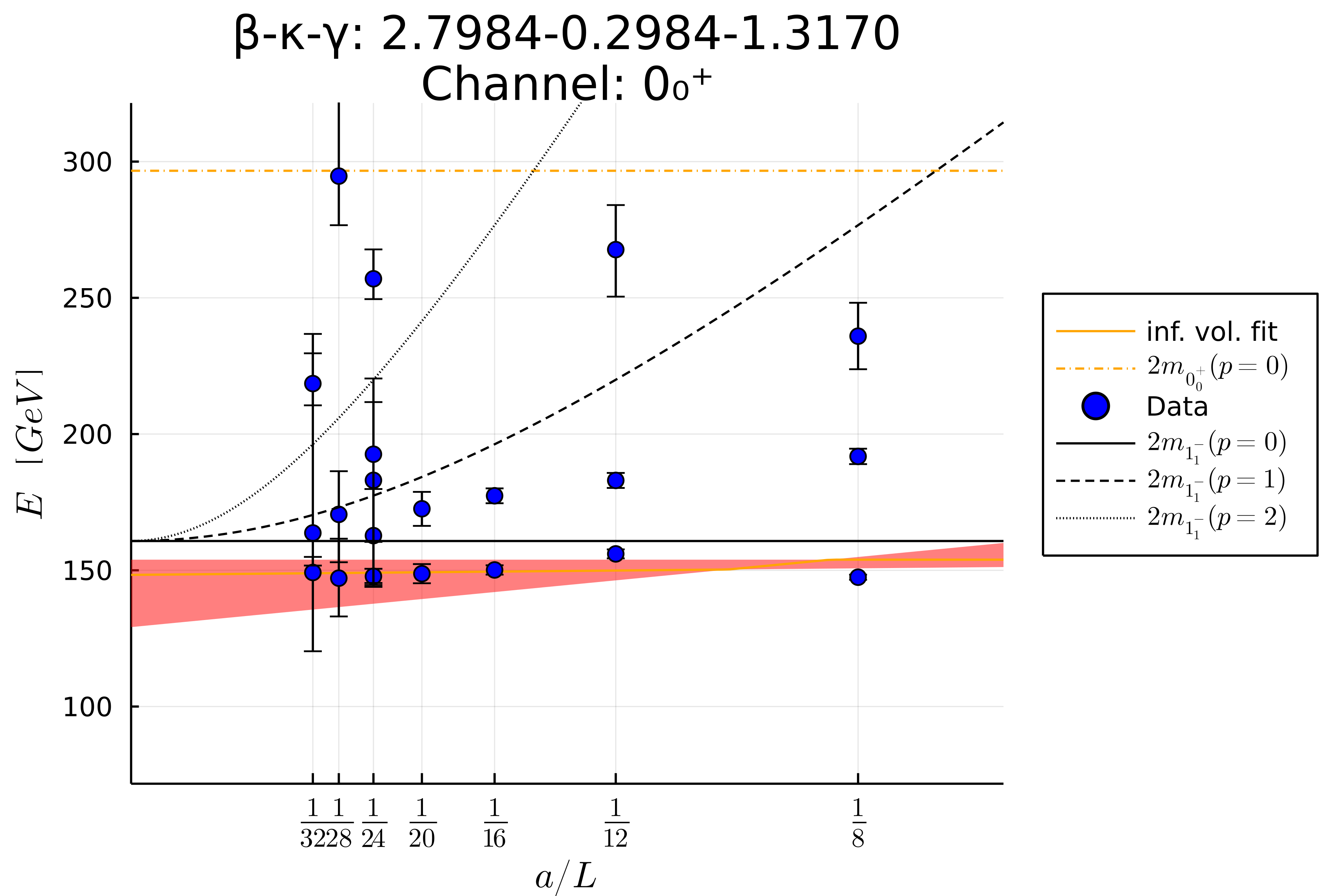}
    \label{fig:spectrum_1}}
    \hfill
    \subfloat[Set2]{\includegraphics[width=0.971\columnwidth]{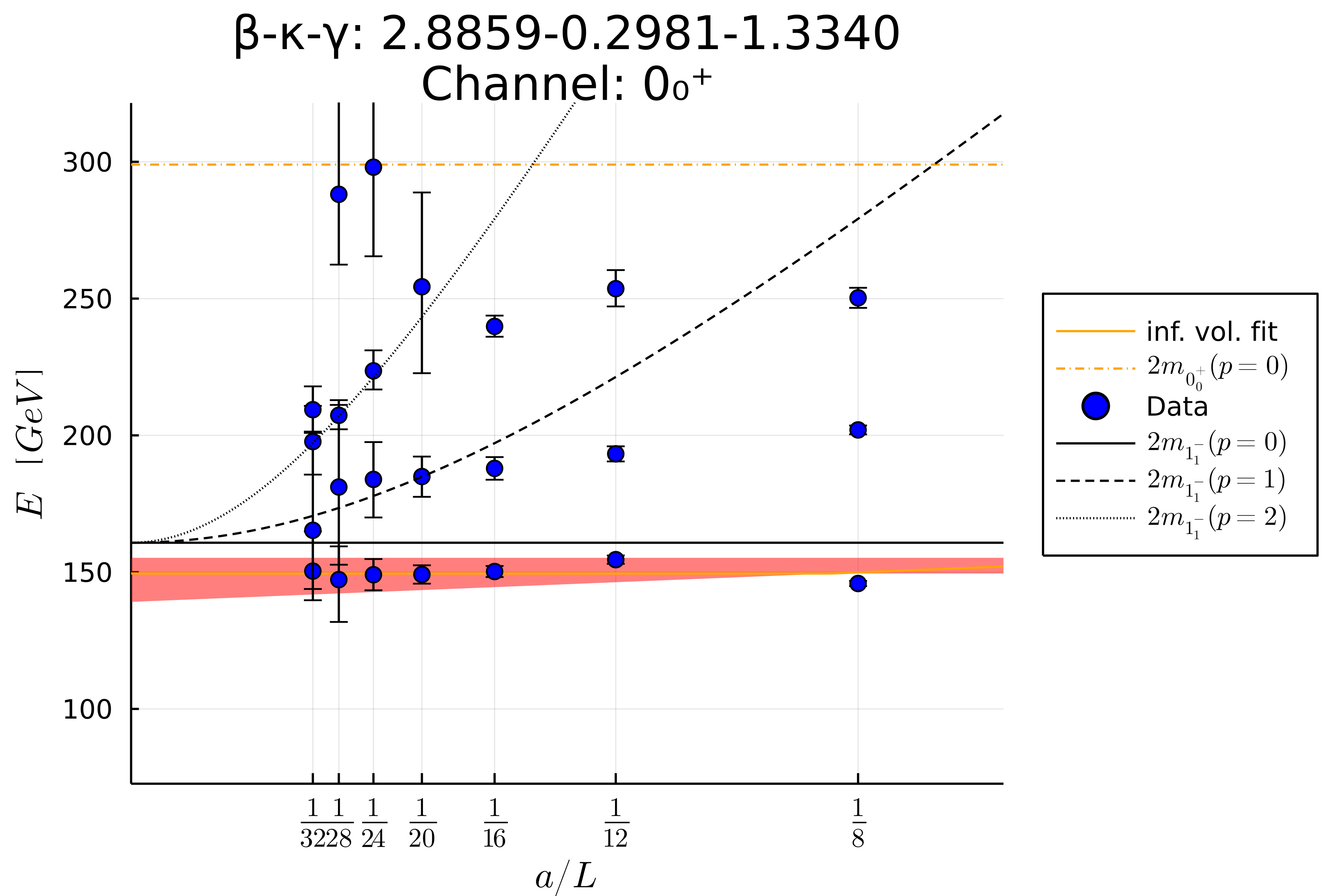}
    \label{fig:spectrum_2}}
    \\
    \subfloat[Set 3]{\includegraphics[width=0.971\columnwidth]{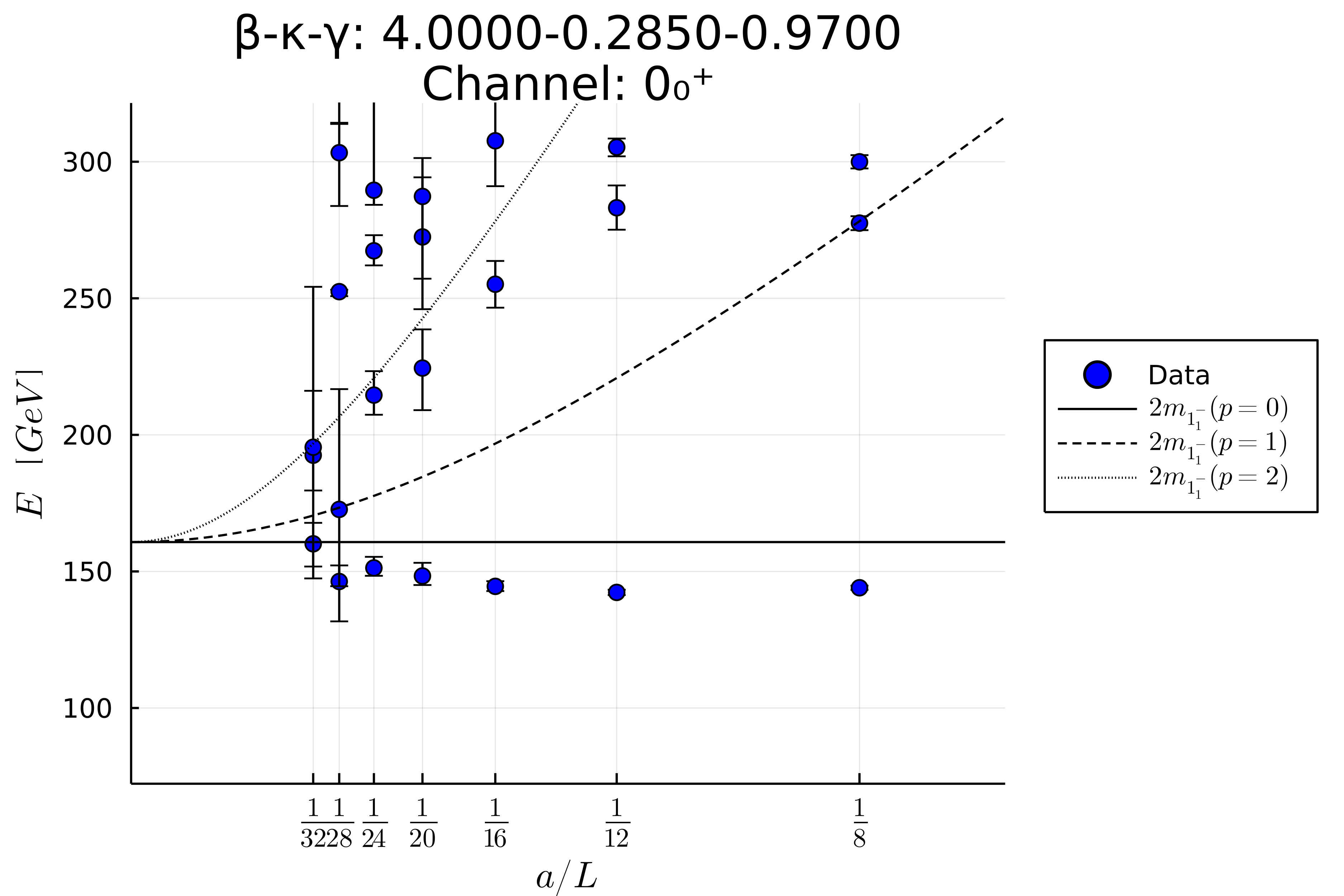}
    \label{fig:spectrum_3}}
    \hfill
    \subfloat[Set 4]{\includegraphics[width=0.971\columnwidth]{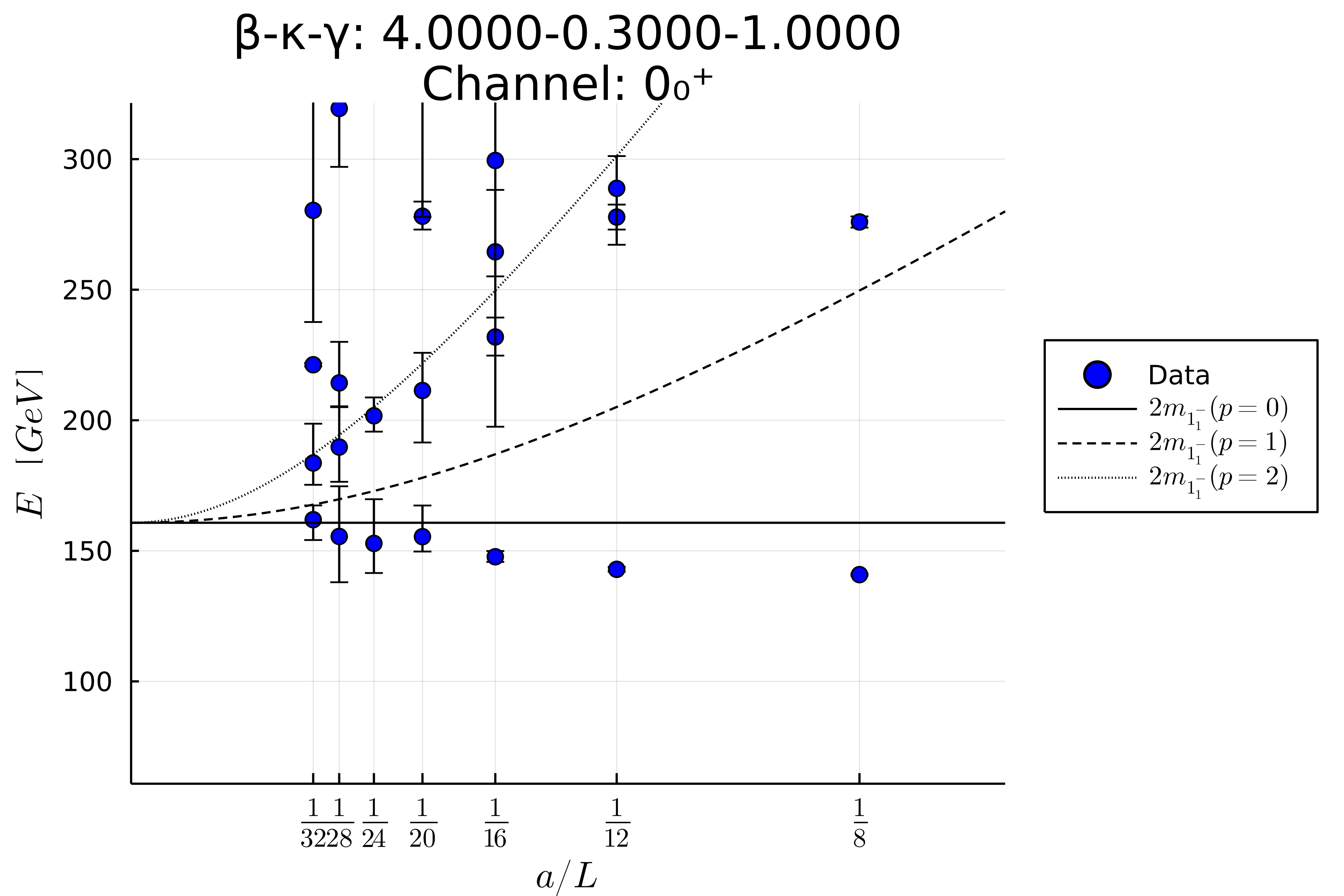}
    \label{fig:spectrum_4}}
    \caption{Energy spectra for the parameter sets in \cref{tab:parameters} with expected non-interacting states. Groundstate fits have been added when suitable (see text). Points that are compatible with \SI{0}{\GeV} or have relative errors larger than \SI{15}{\percent} have been omitted. Full data see \crefrange{tab:data_set1}{tab:data_set4}.}
    \label{fig:spectra}
\end{figure*}

The results\footnotemark for the energy levels in and around the elastic region are shown for the four lattice setups from \cref{tab:parameters} in \cref{fig:spectra} as a function of volume. Only in Sets 1 and 2, \cref{fig:spectrum_1,fig:spectrum_2} respectively, we find a state below the threshold as well as a state which evolves with volume towards the elastic threshold. Thus, we interpret the results in both cases as having a genuine and stable bound state in the $0_0^+$ channel, which would act as a physical version of the Higgs. The masses are given by $148^{+6}_{-20}\si{\GeV}$ and $149^{+6}_{-10}\si{\GeV}$ in the infinite-volume limit, respectively. The large uncertainties of these quantities are mainly due to the growing errors for larger lattices and the corresponding extrapolation. Thus it is very likely that the infinite volume errors are overestimated. However one important feature is that these states are not consistent with the elastic threshold at \SI{160.75}{\GeV}. Further the phase shift analysis below will confirm this interpretation. Thus the inelastic threshold is set from the mean values to \SI{296}{\GeV} and \SI{298}{\GeV}, respectively, rather than at the 4 $1_1^-$ threshold at \SI{321.5}{\GeV}, though the difference is small. This is indicated in \cref{fig:spectrum_1,fig:spectrum_2} by the dashed orange line, which is beneath the top of the plots.

In the other two sets we do not see a second state at or below the elastic threshold. We therefore interpret the single state as the elastic threshold scattering state. Thus, if a $0_0^+$ state should exist, it does so only in the form of a resonance above the elastic threshold, i. e.\ with a mass of at least \SI{160.75}{\GeV}.

\footnotetext[\value{footnote}]{All figures except for \cref{fig:diffxs} have been created using \cite{Julia,*JuliaPlots}. For involved analytic calculations and \cref{fig:diffxs} \cite{Mathematica} has been used.}

We see that in all cases within the elastic region there are substantial differences to the non-interacting scattering states\footnote{This is substantially different from the results in \cite{Wurtz:2013ova}, where heavily oversmeared basic operators. This tends to erase such deviations, and a much smaller bare gauge coupling was employed.}. In addition, we can also infer that not for all sets and volumes we were able to identify all scattering states. Also, many have substantial errors, despite the employed statistics of order $10^5$ decorrelated configurations for every volume and set.

\subsection{\label{subsec:phase shift}Phase shifts and resonance parameters}

\begin{figure*}[!t]
    \centering
    \subfloat[Set 1]{
        \includegraphics[width=0.971\columnwidth]{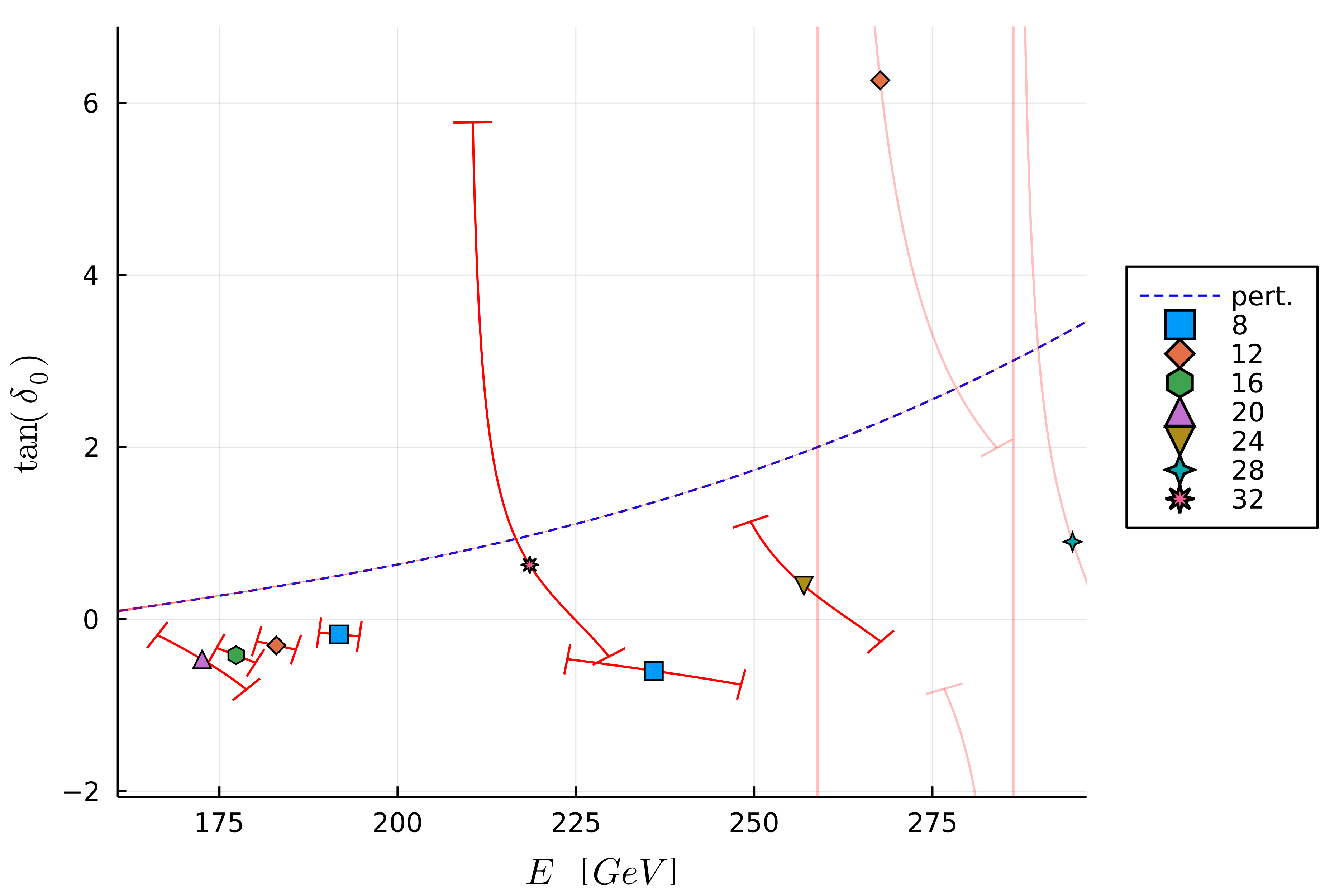}
        \label{fig:tan_1}
    }
    \hfill
    \subfloat[Set 2]{
        \includegraphics[width=0.971\columnwidth]{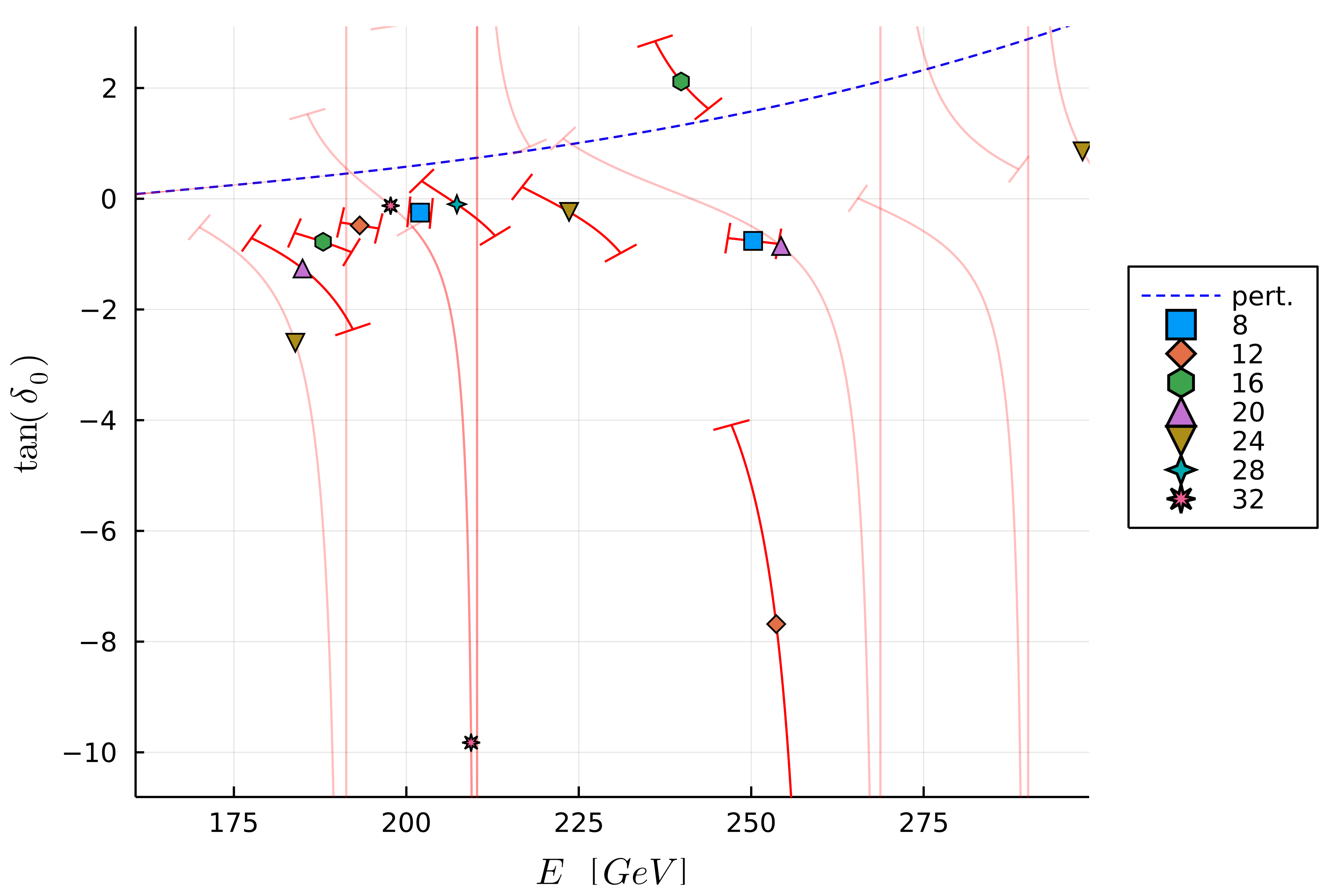}
        \label{fig:tan_2}
    }
    \\
    \subfloat[Set 3]{
        \includegraphics[width=0.971\columnwidth]{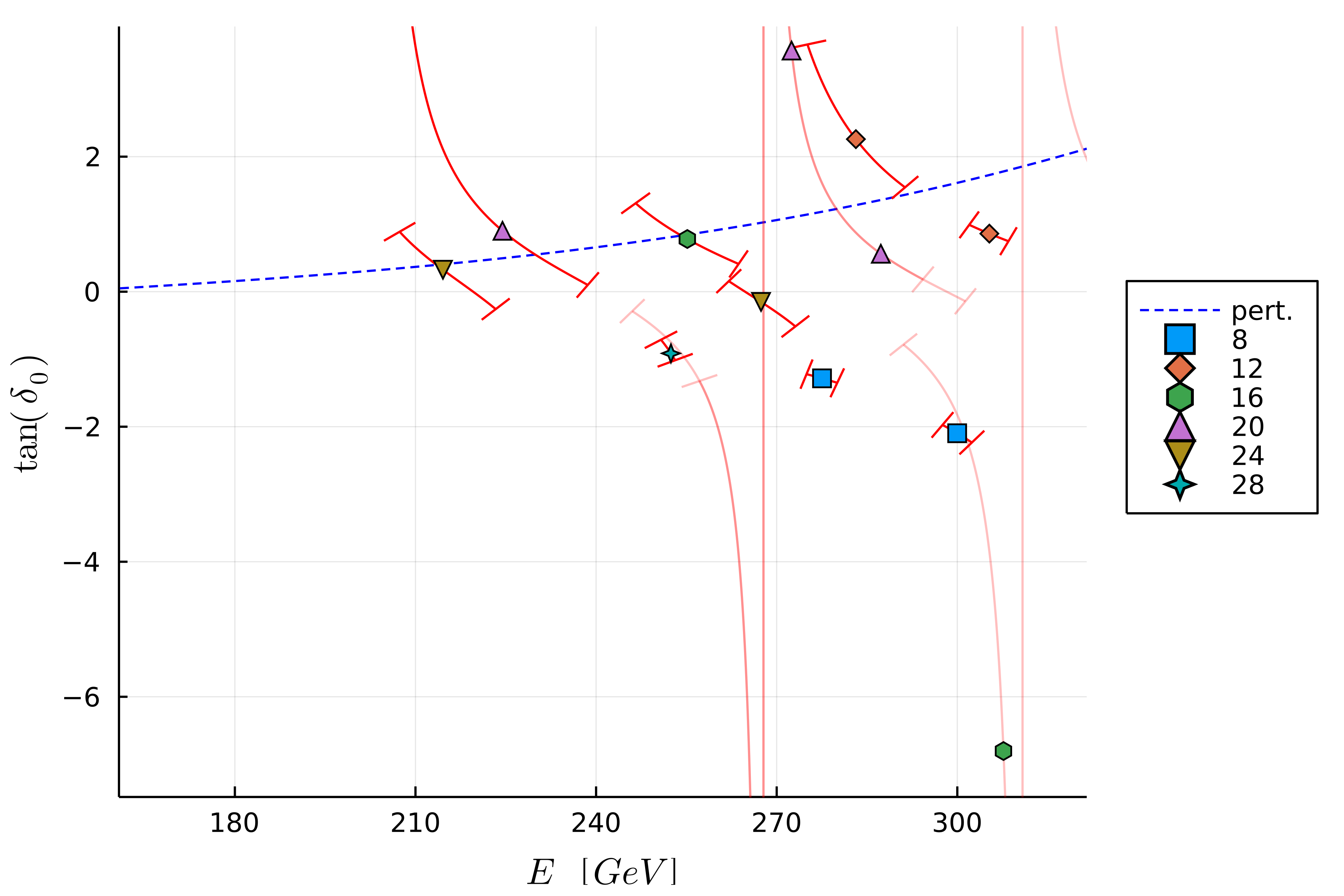}
        \label{fig:tan_3}
    }
    \hfill
    \subfloat[Set 4]{
        \includegraphics[width=0.971\columnwidth]{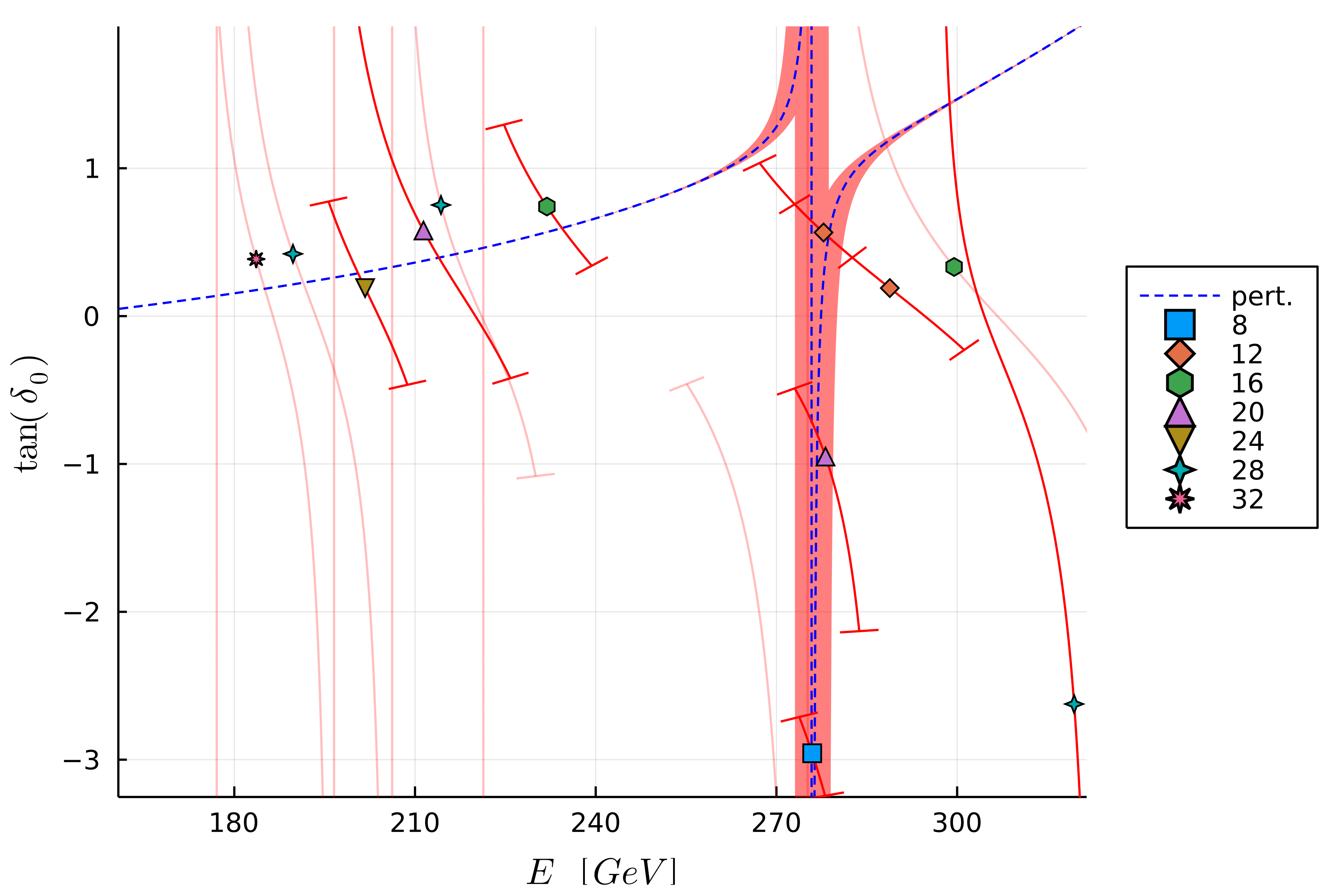}
        \label{fig:tan_4}
    }
    \caption{The (tangent of the) phase shift as a function of energy for the parameter sets 1-4 in (a)-(d) as listed in \cref{tab:parameters}. Points with errors crossing one singularity are shown fainter while those crossing multiple singularities have been dropped. In (d) one data point at \SI{221}{\GeV} lies outside the plot pane, but respective errors are still shown.}
    \label{fig:tans}
\end{figure*}

\begin{figure*}[!t]
    \centering
    \subfloat[Set 1]{
        \includegraphics[width=0.971\columnwidth]{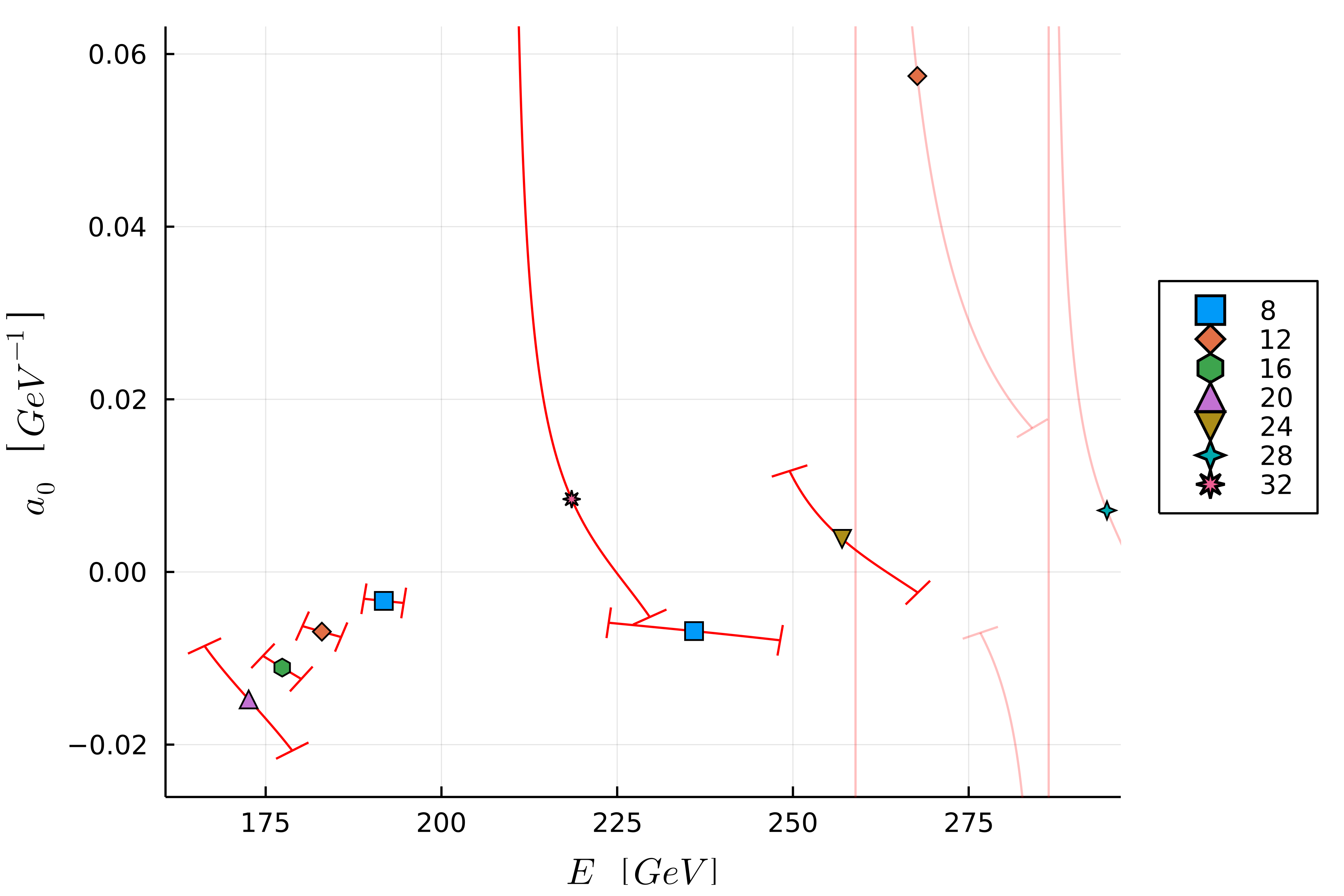}
        \label{fig:scatlen_1}
    }
    \hfill
    \subfloat[Set 2]{
        \includegraphics[width=0.971\columnwidth]{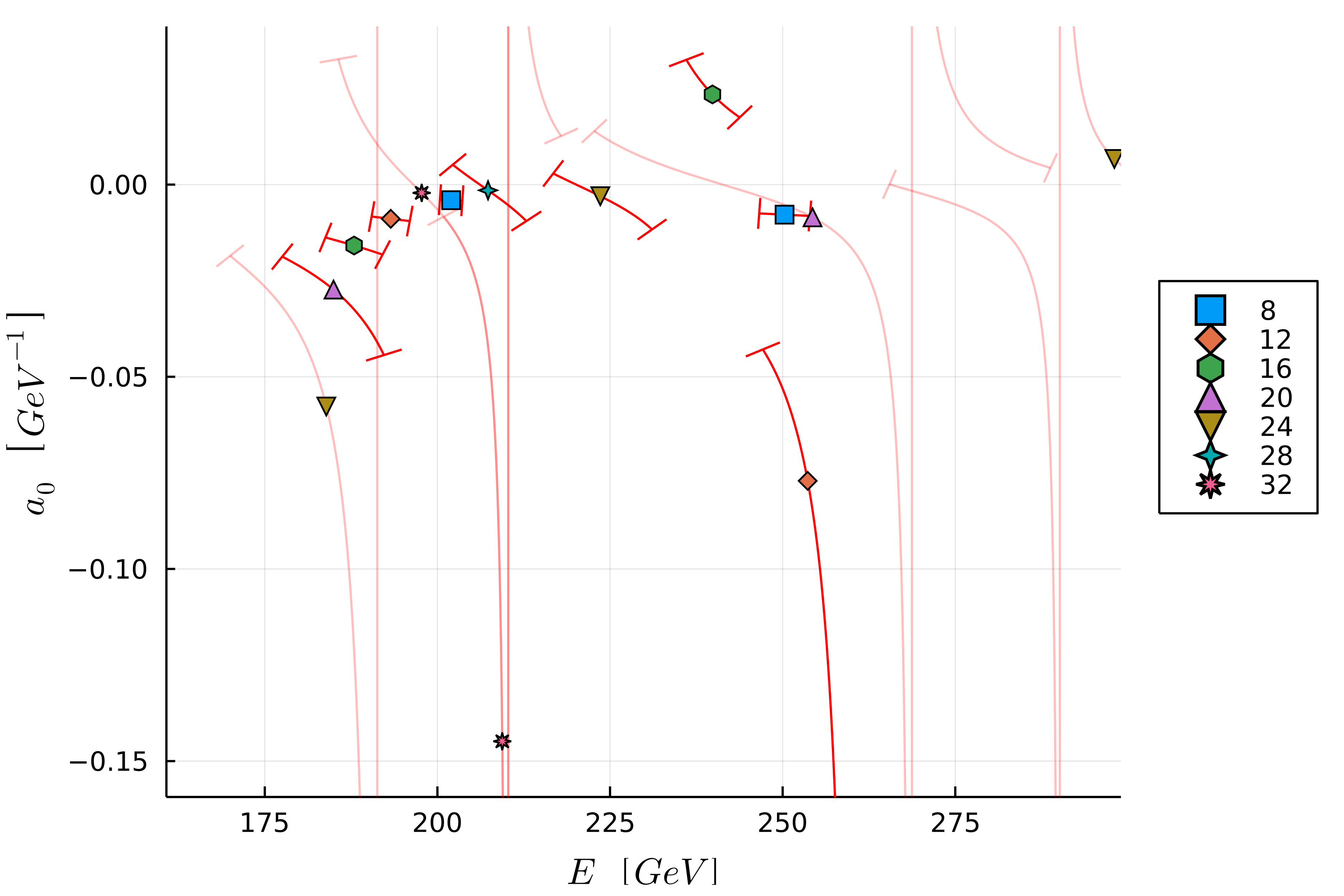}
        \label{fig:scatlen_2}
    }
    \\
    \subfloat[Set 3]{
        \includegraphics[width=0.971\columnwidth]{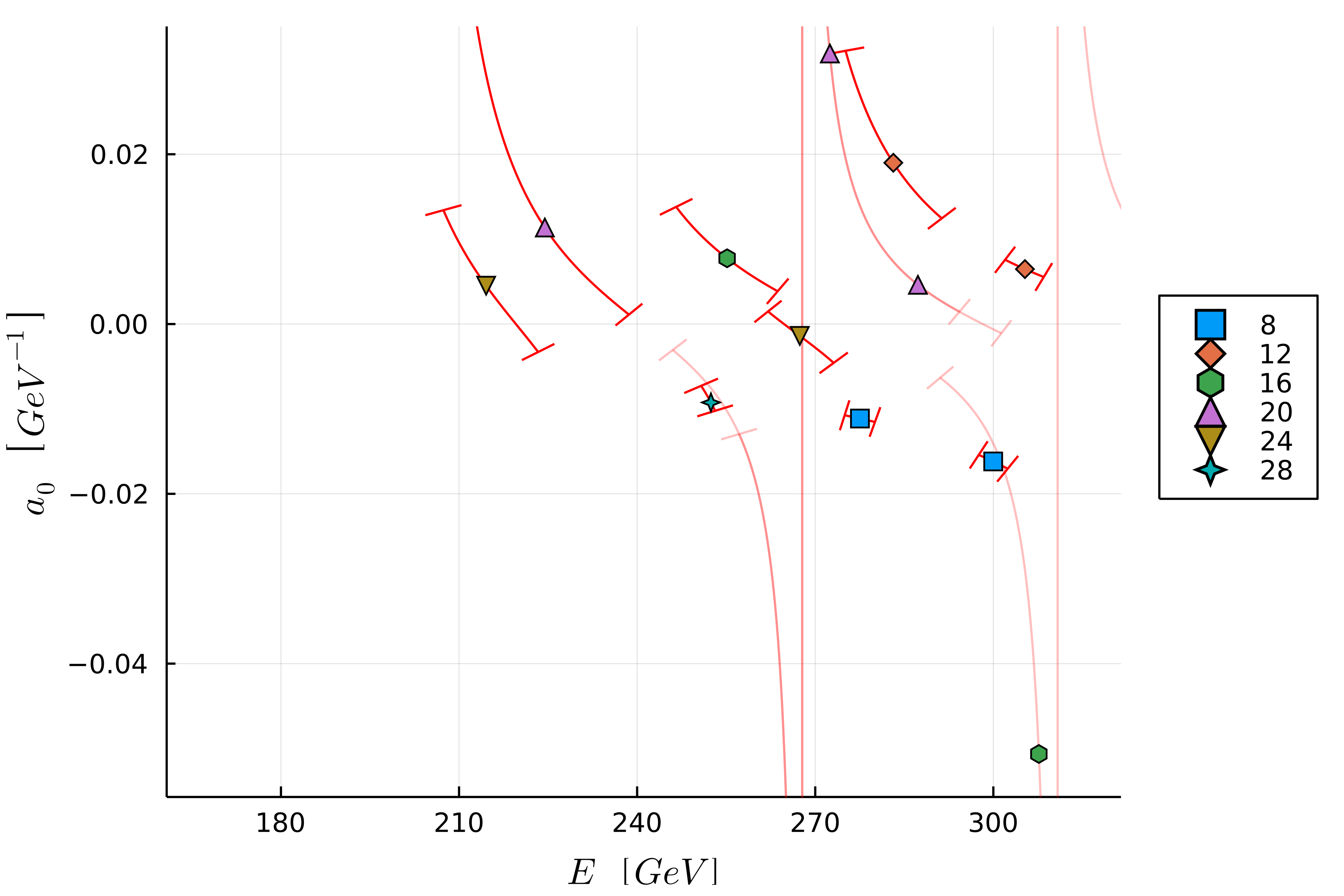}
        \label{fig:scatlen_3}
    }
    \hfill
    \subfloat[Set 4]{
        \includegraphics[width=0.971\columnwidth]{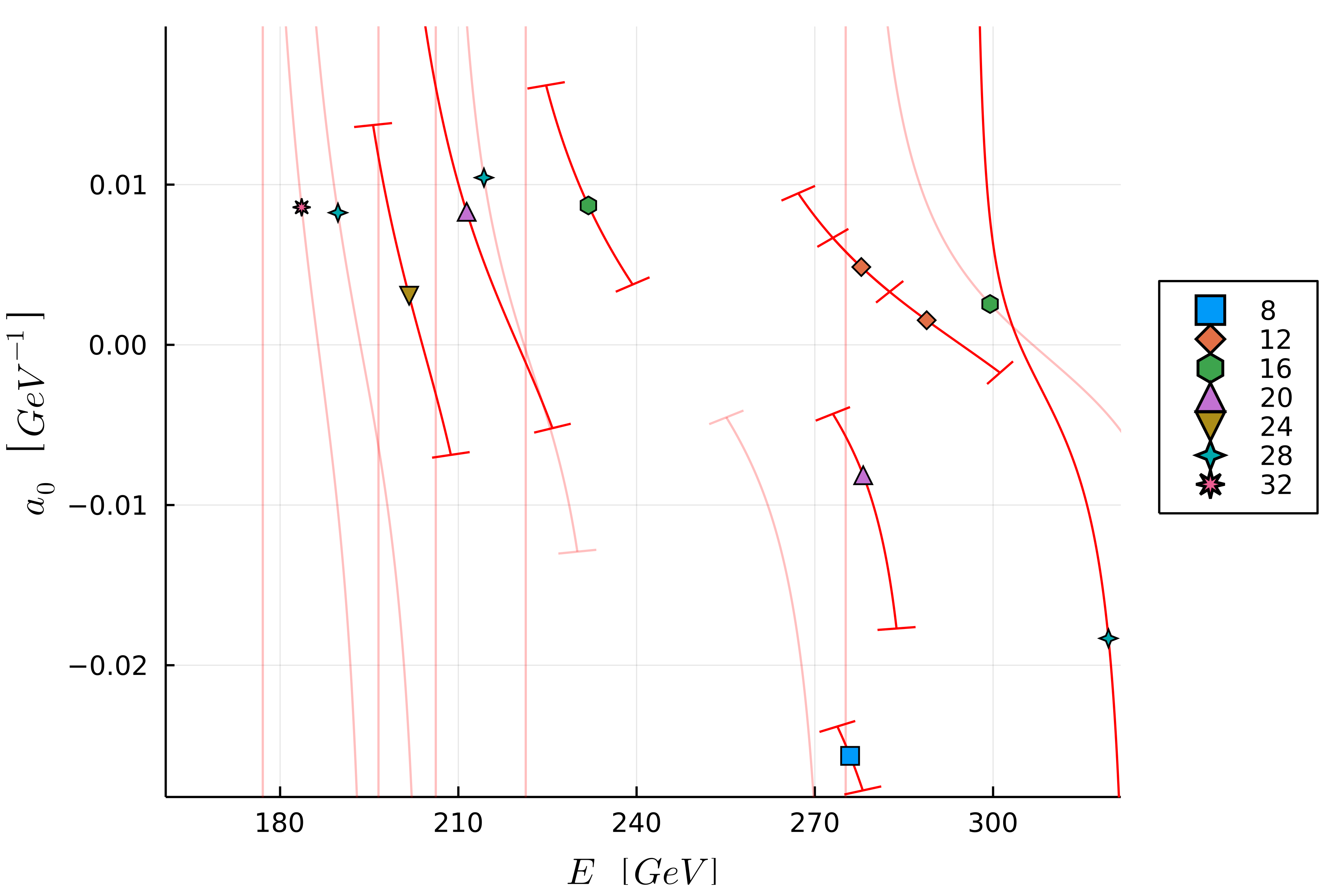}
        \label{fig:scatlen_4}
    }
    \caption{The effective scattering length for the parameter sets 1–4 in (a)–(d) as listed in \cref{tab:parameters}. This quantity becomes the actual scattering length at the inelastic threshold. Points with errors crossing one singularity are shown fainter while those crossing multiple singularities have been dropped. In (c) one data point at \SI{269}{\GeV} lies outside the plot pane, but respective errors are still shown.}
    \label{fig:scats}
\end{figure*}

Our results for the (tangent of the) phase shifts are plotted in figures \crefrange{fig:tan_1}{fig:tan_4}, also in comparison to the perturbative predictions from section \ref{sec:pt}. We also plot $\tan(\delta_0)/|\vec p|$ in figures \crefrange{fig:scatlen_1}{fig:scatlen_4}, which yields the scattering length at threshold. Note that the perturbative prediction for the tangens of the phase shift has always positive value and thus the scattering length is also always positive.

The simplest picture arises for set 3 \cref{fig:tan_3,fig:scatlen_3}. In this case there is no bound state below threshold, and no recognizable feature indicating a resonance. Thus we use the analytic predictions for a Higgs mass much larger than all other scales (see \cref{eqn:born_woh}). Therefore the analytic predictions do not contain any uncertainties. Our data scatter then around the so obtained perturbative prediction. Also the scattering length is within errors rather consistent with zero, as in the perturbative case. We interpret this result therefore such that there is no light scalar resonance, i. e.\ with a mass below the inelastic threshold of \SI{321.5}{\GeV}, in this channel. The interaction is essentially weak, and acceptably described by the analytic prediction.

The situation for sets 1 and 2, \cref{fig:tan_1,fig:scatlen_1} and \cref{fig:tan_2,fig:scatlen_2} respectively, is relatively similar. Here the perturbative prediction is entirely fixed by the properties of the bound states and the running coupling. The error of the perturbative prediction is thus entirely due to the uncertainty of the bound state mass shown in \cref{tab:parameters}. Although the uncertainty of the mass is non-negligible it does not have a remarkable influence on the predicted functional form. The uncertainties are shown as a red ribbon in \cref{fig:tan_1,fig:tan_2} respectively, which however can only be seen close to the threshold. Away from the threshold the data again scatter around the analytic prediction, which therefore gives an acceptable description of it. However, when approaching the elastic threshold the lattice data significantly and consistently starts to deviate from the analytic prediction, and becomes substantially negative. As noted, this is indicative, and thereby consistent, with a bound state below threshold, as was also identified in the spectroscopical analysis. While a simple extrapolation to threshold is difficult especially for set 2, the results are broadly consistent with an inverse scattering length of order \SI{40}{\GeV}. This will be further discussed in \cref{sec:exp}.

However, already at this stage this is a remarkable result, as this the same ballpark as was obtained for the custodial/weak radius of the $1^-_1$ state in \cite{Maas:2018ska}. As both bound states emerge from the same weak interactions, a rough agreement of their typical size parameters is encouraging. This also emphasizes again the interpretation in \cite{Maas:2018ska} that the bound states in the weak-Higgs sector exhibit a deep-inelastic-scattering-like (DIS-like) behavior: When probed at low energies, they show a coherent reaction of the whole bound state, here demonstrated by a typical bound-state behavior of the scattering length close to the elastic threshold. Probing at high energies, here towards the inelastic threshold, the reaction is that of the FMS-dominating individual constituent, and thus as obtained in the perturbative approach.

Whether this size parameter is a genuine non-perturbative feature, or whether it can be captured by the FMS mechanism beyond leading order in the vev is an open question, and will have to await detailed calculations in the future.

Set 4 finally shows a third behavior. The data is again relatively well described by the perturbative expression. Especially, close to the elastic threshold we do not see any indication of a sizable negative scattering length, consistent with the absence of a bound state below threshold in the spectroscopical analysis.

However, around \SI{275}{\GeV} we observed a strong deviation of the statistical significant data, tending towards very large negative values of $\tan(\delta_0)$. Such a large deviation is indicative of a resonance \cite{Briceno:2017max,Luscher:1990ux}. However, attempting a usual Breit-Wigner fit is not successful, mainly due to the still relatively poor quality of the data. But using the mass of the Higgs in the perturbative description as a free parameter we find a quite good fit of the data for a Higgs mass of about  $275(3)\si{\GeV}$. The uncertainty of the fit here is determined by error-propagation using the Jacobian of the model function. Given the otherwise good agreement, this strongly suggests the existence of a resonance in this channel at this energy.

Using the same Born-level-type calculation would yield a width of the bound state at this energy of about \SI{40}{\GeV} \cite{Bohm:2001yx}. However, the matrix element cannot easily be supplemented by such a width due to the unitarity violation problems of the Born amplitude. Still, a \SI{40}{\GeV} window around the resonance position is still broadly within the errors of the data, and thus it would be possible that a more elaborate analysis could fix this width. In the present system, this is therefore an alternative to the usual Breit-Wigner fit or more sophisticated approaches, which is possible only due to the FMS mechanism.

\subsection{\label{sec:exp}Impact on cross sections}

\begin{figure*}[t!]
    \centering
    \subfloat[Set 1]{
        \includegraphics[width=0.971\columnwidth]{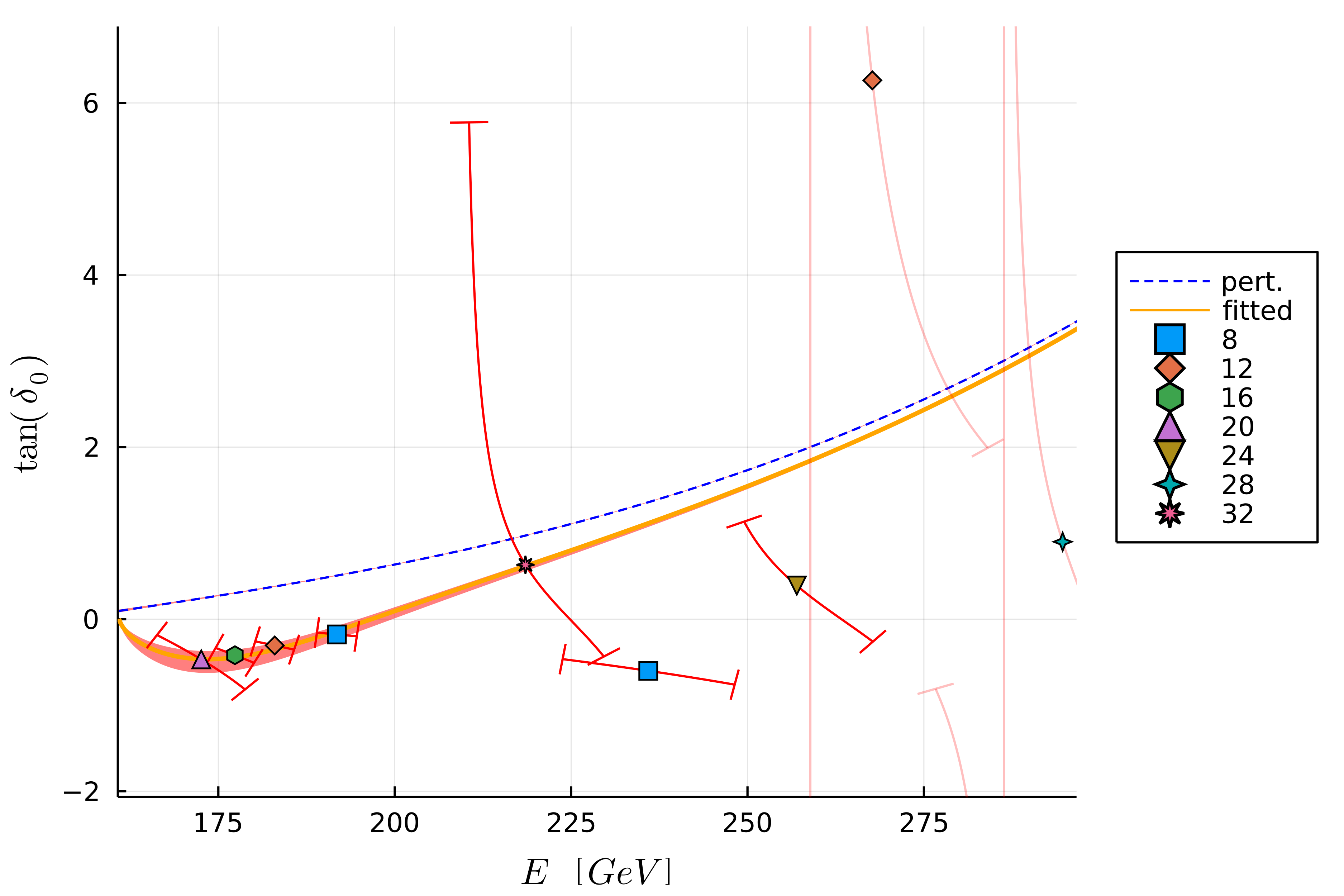}
        \label{fig:fittan_1}
    }
    \hfill
    \subfloat[Set 2]{
        \includegraphics[width=0.971\columnwidth]{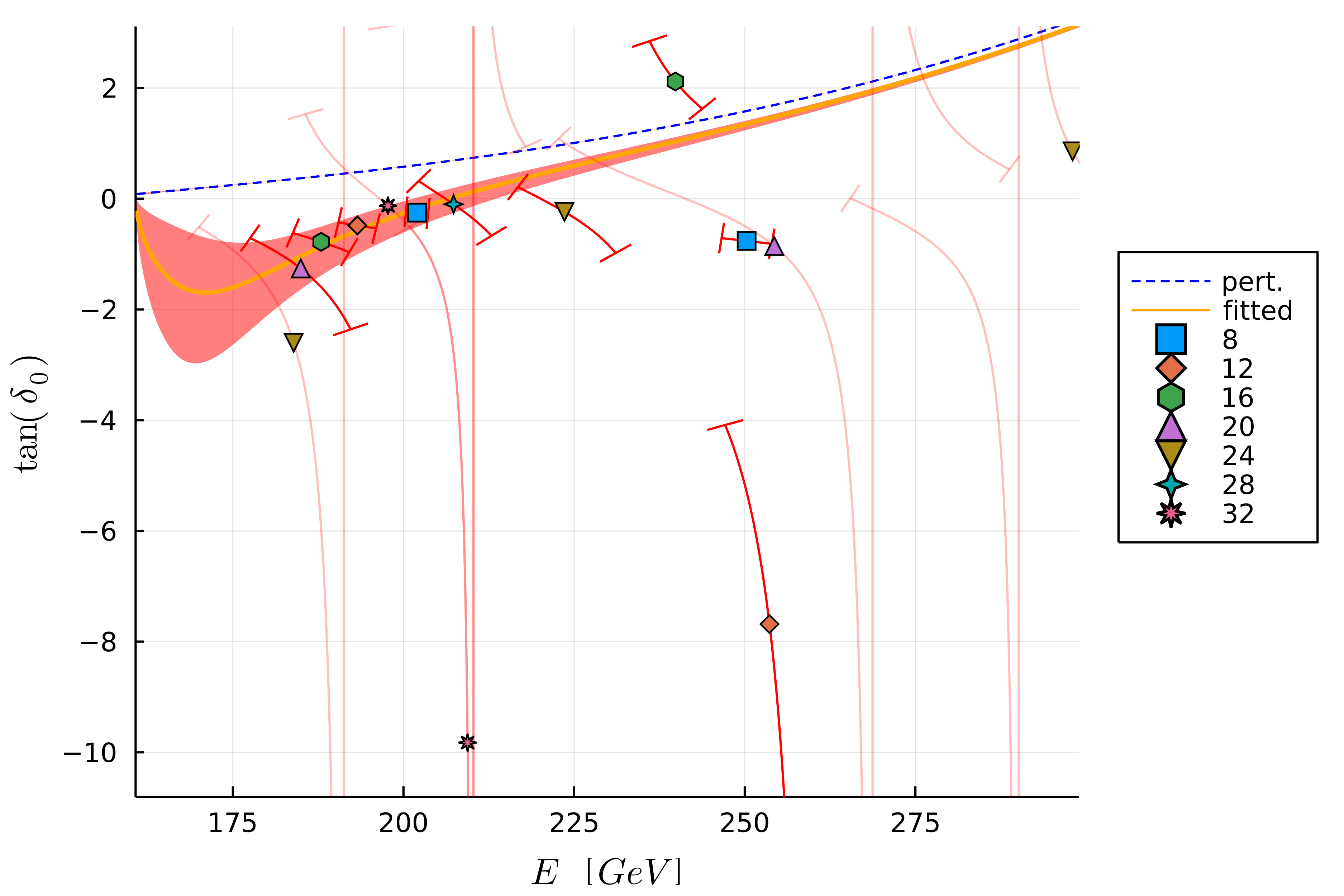}
        \label{fig:fittan_2}
    }
    \caption{The (tangent of the) phase shift in comparison to the fit \cref{eqn:mod_tan} for sets 1 and 2.}
    \label{fig:fittans}
\end{figure*}

Since VBS is used as a primary tool for searching for new physics \cite{Denner:2022pwc,Gallinaro:2020cte}, it is interesting to ask how the observed effects could influence this. Since effects above the inelastic threshold are inaccessible, we focus here on the negative scattering length close to threshold. Since such a negative scattering length would necessarily arise in any scenario with the Higgs being a bound state, e.g.\ composite Higgs scenarios and technicolor \cite{Cacciapaglia:2020kgq}, the observed effect here is indeed yet unaccounted-for SM background\footnote{Of course, our reduced SM setup cannot give a true quantitative statement. Also, in the full SM the Higgs is unstable, though relatively long-lived. Hence, at least a coupled-channel analysis would be needed to repeat this in the full SM \cite{Briceno:2017max}. It remains that there is an attractive component in contrast to the purely repulsive perturbative one.} for such a signal.

VBS with on-shell vector bosons in the initial or final state, and thus in the relevant energy range, is only possible currently at the LHC. ATLAS and CMS so far determine corresponding cross sections, and will determine doubly differential cross sections with (likely) fine enough binning in the future \cite{Covarelli:2021fra,Covarelli:2021gyz}. The relevant question is thus what the impact on these cross sections are.

The deviation in $\tan\qty(\delta)$ visible for sets 1 and 2 in \cref{fig:tans} can be well accounted for by the replacement
\begin{equation}\label{eqn:mod_tan}
    \tan\qty(\delta_\text{Born})\to \tan\qty(\delta_\text{Born}) - 2\Delta f
\end{equation}
where the factor 2 has been chosen for later convenience. The additional contribution $\Delta f$ is of the form
\begin{equation}\label{eqn:deltaf}
    \Delta f(E)=\frac{\sqrt{E^2-4m_W^2}}{4\left(a_0^{-1}+\left(\frac{\sqrt{E^2-4m_W^2}}{a_1^{-1}}\right)^4\right)}.
\end{equation}
with $a_0^{-1}$ and $a_1^{-1}$ being some coefficients used to model the influence of the bound-state structure of the Higgs. Because this contribution will only appear in the channel with the Higgs, the $s$-wave, it is independent of the scattering angle. Also, since the $\tan\qty(\delta_\text{Born})$ is always positive for $m_H\lessapprox 2m_W$ and becomes negligible near the threshold, we immediately see that $a_0^{-1}$ is indeed the scattering length. Thus this will give an estimate on the scattering length that has not been possible by simple extrapolation of the data, as discussed in the previous chapter.
Fitting the data in \cref{fig:tans} with the modified relation in \cref{eqn:mod_tan} yields $a_0^{-1}$ and $a_1^{-1}$ being roughly $39^{+10}_{-6}\si{\GeV}$ and $41^{+2}_{-1}\si{\GeV}$ for set 1 and $12^{+10}_{-8}\si{\GeV}$ and $43^{+5}_{-3}\si{\GeV}$ for set 2, respectively. This agrees with the rough estimate of \SI{40}{\GeV} from before.

The result of including \cref{eqn:mod_tan} is shown in \cref{fig:fittan_1} and \cref{fig:fittan_2} for sets 1 and 2, respectively. While the largest modification is close to threshold, this reduces the phase shift also at larger momenta, bringing it overall into better agreement with the data. The uncertainty is dominated by the scattering length, especially for set 2.

The corresponding matrix element is then
\begin{equation}
    \mathcal{M}=\mathcal{M}_\text{Born}-32\pi \Delta f,
\end{equation}
because at this level of reunitarization the partial wave amplitude is just $\tan\qty(\delta)$. Because the process is elastic, the energies and total values of three momenta of the initial states and final states are equal, and thus the differential cross section is given by \cite{Bohm:2001yx}
\begin{align}
    \frac{d\sigma}{d\Omega}&=\frac{1}{64\pi^2 E^2}\left|\mathcal{M}\right|^2\nonumber\\&=\frac{1}{64\pi^2 E^2}\left(\mathcal{M}_\text{Born}^2-64\pi \Delta f\mathcal{M}_\text{Born}+\qty(32\pi\Delta f)^2\right)
\end{align}
because both $\mathcal{M}_\text{Born}$ and $\Delta f$ are real. Normalizing this to the differential Born level cross section yields
\begin{equation}\label{eqn:diffxs}
    \left.\qty(\frac{d\sigma}{d\Omega})\middle/\qty(\frac{d\sigma}{d\Omega})_{\text{Born}}\right.=1-\frac{64\pi \Delta f}{\mathcal{M}_\text{Born}}\left(1-\frac{16\pi \Delta f}{\mathcal{M}_\text{Born}}\right)
\end{equation}
which is thus close to threshold a sizable contribution, but quickly diminishes at higher energies.

\begin{figure*}[t!]
    \centering
    \subfloat[Set 1]{
        \includegraphics[width=0.971\columnwidth]{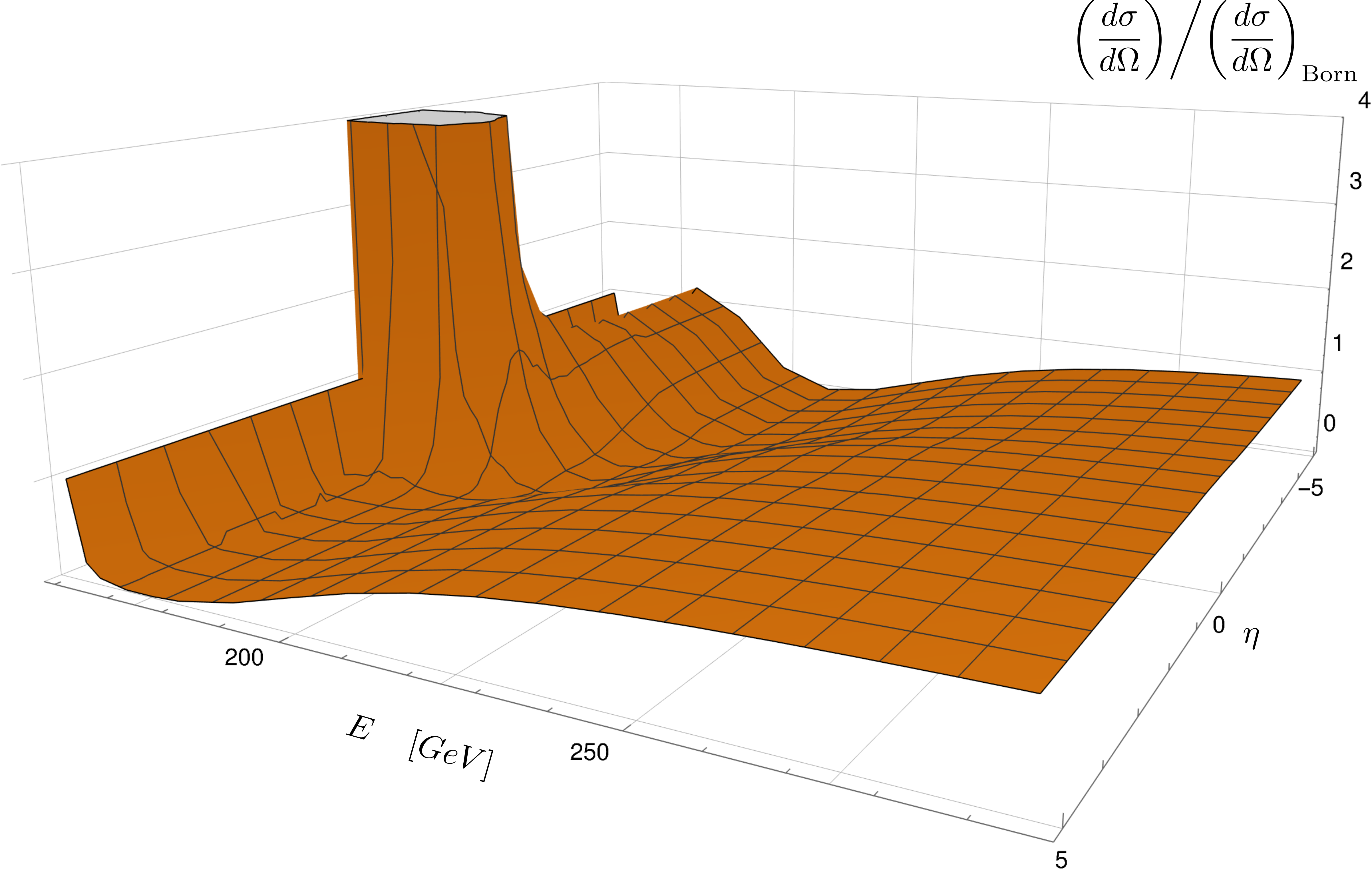}
        \label{fig:diffxs_1}
    }
    \hfill
    \subfloat[Set 2]{
        \includegraphics[width=0.971\columnwidth]{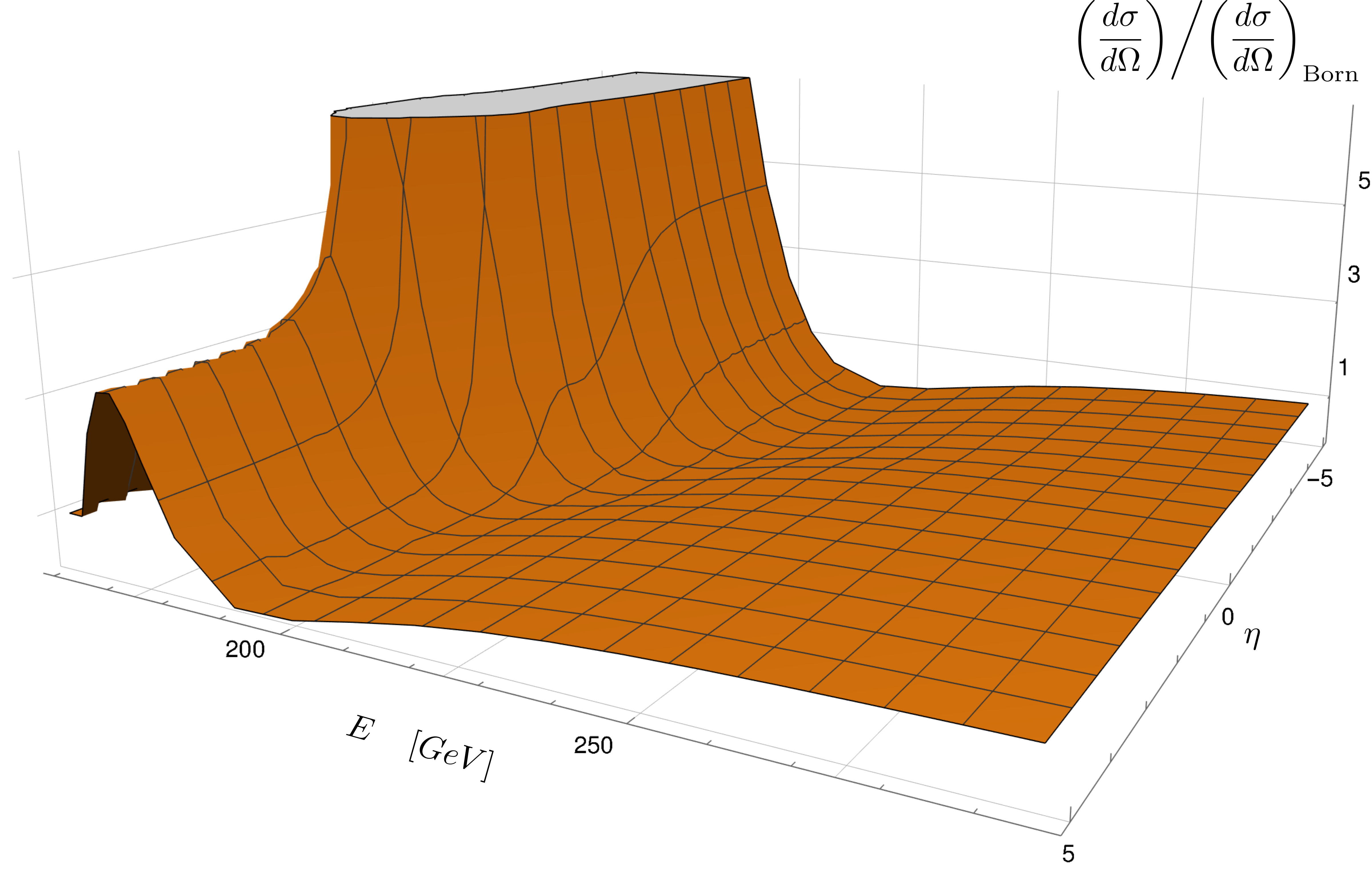}
        \label{fig:diffxs_2}
    }
    \caption{Full normalized differential cross section as given in \cref{eqn:diffxs} as a function of energy and pseudo-rapidity $\eta$. Errors of the parameters  $m_H$, $a_0^{-1}$ and $a_1^{-1}$ have been omitted for obtaining these illustrations.}
    \label{fig:diffxs}
\end{figure*}

\begin{figure*}[t!]
    \centering
    \subfloat[Set 1]{
        \includegraphics[width=0.971\columnwidth]{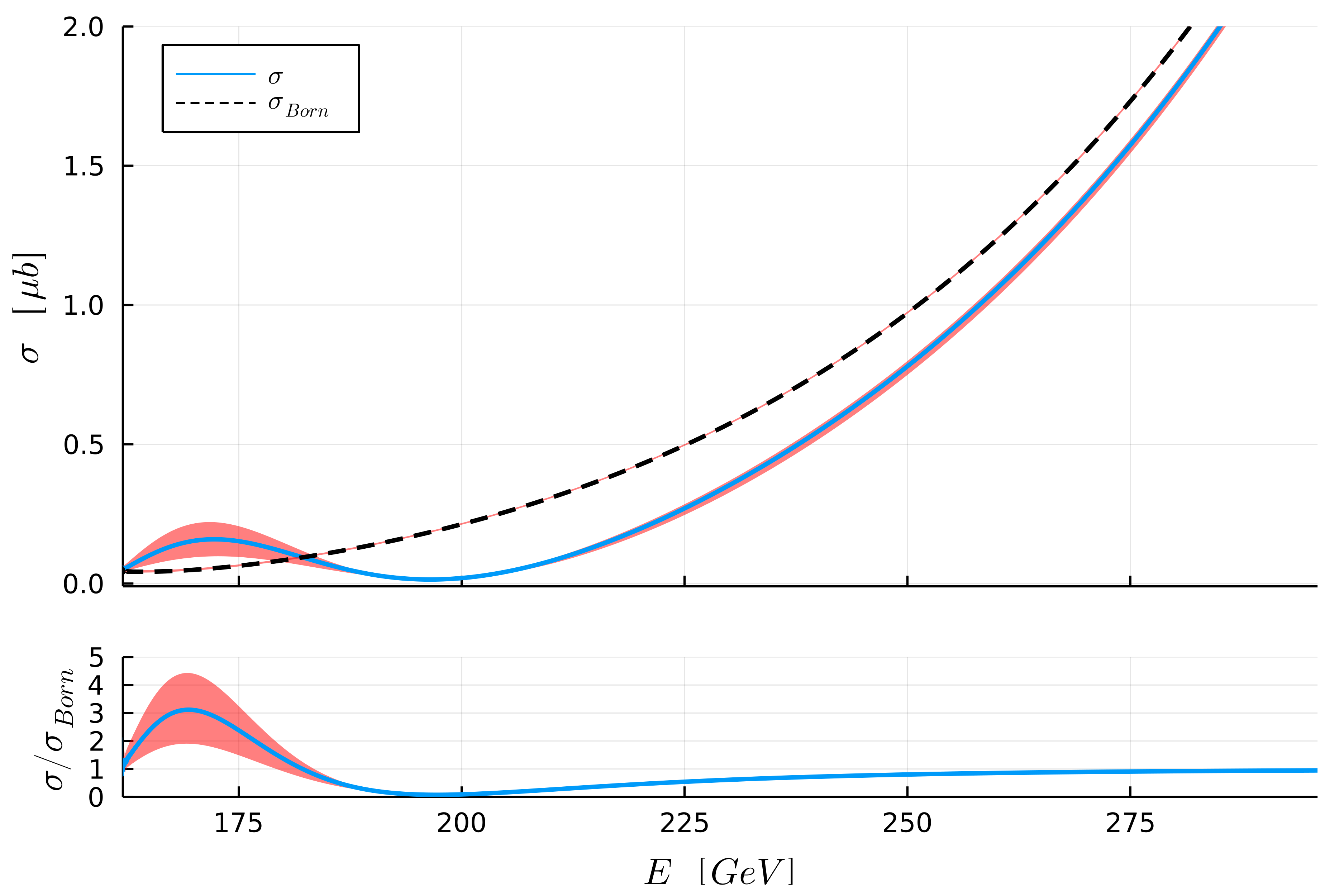}
        \label{fig:xs_1}
    }
    \hfill
    \subfloat[Set 2]{
        \includegraphics[width=0.971\columnwidth]{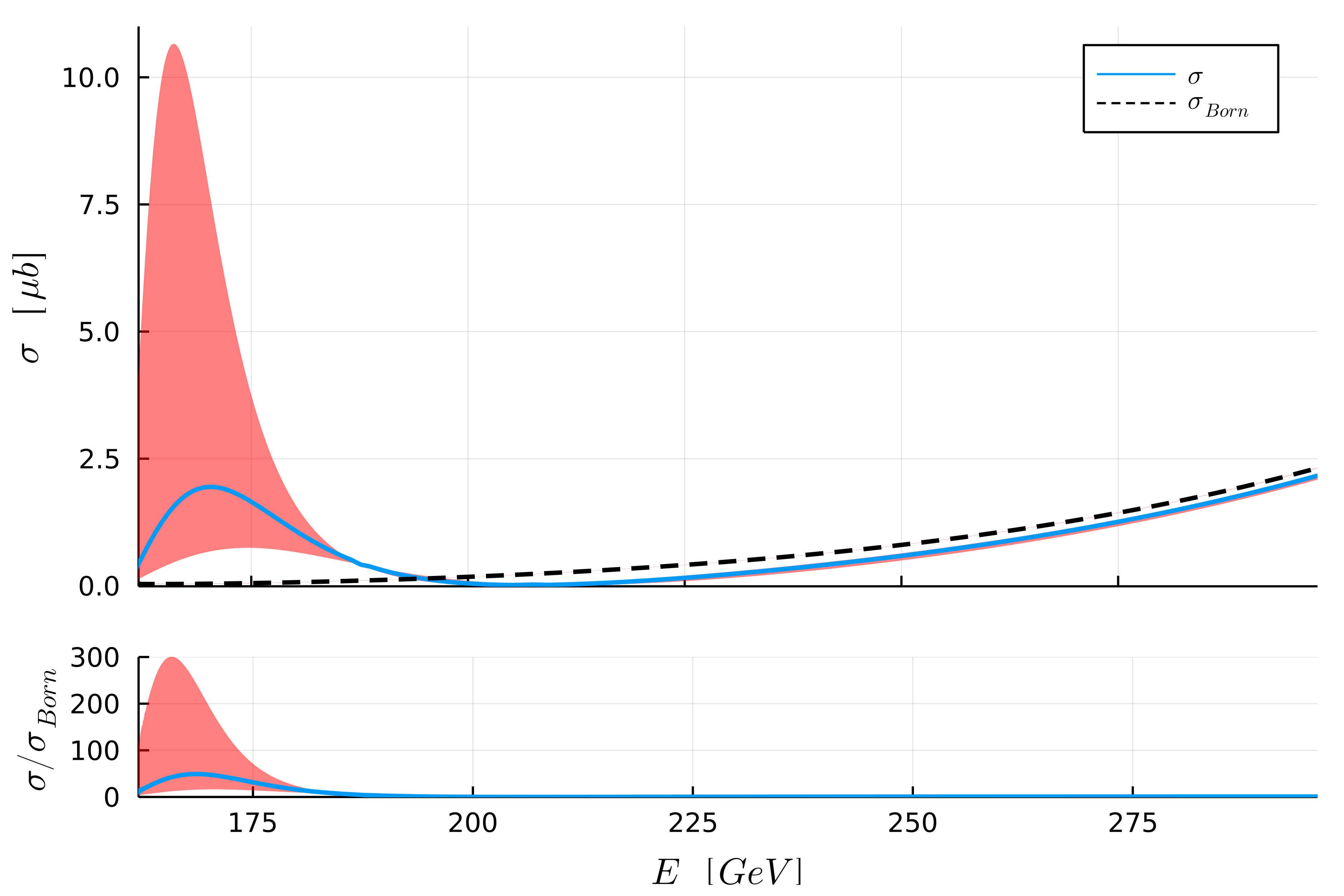}
        \label{fig:xs_2}
    }
    \caption{Comparison (upper) and ratio (lower) of integrated born cross section and the full cross section in microbarn ($\mu b$) as given in \cref{eqn:xs} as a function of energy.}
    \label{fig:xs}
\end{figure*}

Integrating the differential cross sections and combining the results yields the modified total cross section
\begin{equation}\label{eqn:xs}
    \sigma = \sigma_\text{Born}-\frac{64\pi\Delta f}{E^2}\qty(\tan\qty(\delta_{\text{Born}})-\Delta f).
\end{equation}
Note that the term in parentheses is not the same as the initial replacement in \cref{eqn:mod_tan}, due to the missing factor of 2. However we still can see that the difference in the cross section is mainly dominated by the deviation in the tangent.

Using the fits, these deviations are shown in \cref{fig:diffxs} for the differential cross section and in \cref{fig:xs} for the integrated cross section. In particular, the largest deviation appears close to the threshold when being scattered perpendicular to the incoming particles. Near the threshold even the total cross section differs by a factor of 4. It is also visible that the slightly different values for $a_0$ and $a_1$ in the two sets give very different effects. In fact, changing either parameter by a factor 2 changes the impact on the cross section by a much larger factor. In fact, reducing $a_0$ and $a_1$ sufficiently leads to a suppression of the integrated cross section for all energies, rather than an enhancement in some kinematic range. The impact is thus strongly dependent on the quantitative numbers. This can also be seen in the integrated cross sections. While both cross sections are closing in to each other in the high energy region, the behavior close to the threshold is qualitatively different. However again the effect is very sensitive to the actual parameters $a_0$ and $a_1$ as can be seen from the error bounds.

Because this is a reduced SM, and especially sets 1 and 2 were at stronger weak coupling, the effect is thus likely substantially overestimated. Also, the large difference between sets 1 and 2 while having very similar input parameters shows the necessity to have very good quantitative control to predict the impact on actual cross section measurements, not to mention the inclusion of background. Hence, the results here should be considered to be given the correct qualitative behavior, but not a quantitative prediction.

It should also be noted that the effect originates from the presence of the Higgs below threshold. VBS with vector bosons with non-zero total weak/custodial charge will therefore not be affected. Furthermore, the mixing between the $Z$ and $\gamma$ will in practice affect even the $ZZ$ channel as this modifies the involved coupling constants and masses.

\section{\label{sec:wrapup}Summary and Conclusion}

We have presented the first fully gauge-invariant study of VBS in a reduced SM setup in the elastic region of the uncharged $J^{P}_\mathcal{C}=0^{+}_0$ partial wave. We confirmed the absence of genuine non-perturbative bound states in the scalar singlet channel below the inelastic threshold, confirming previous results based on level counting \cite{Wurtz:2013ova,Maas:2014pba,Maas:2017wzi}. We also support that an FMS Born-level analysis is able to describe the scattering process quite well, including the appearance of a resonance in the elastic region. Only in presence of a Higgs below threshold the appearing, and expected, negative scattering length is not fully captured at leading order. It remains to be seen whether this can be remedied at higher orders in the FMS expansion. Additionally it should be mentioned that recent investigations showed, that NLO corrections to VBS in the other couplings modifies the resulting total cross section at the LHC by up to \SI{10}{\percent} \cite{Biedermann:2017bss,Denner:2019tmn,Denner:2020zit,Denner:2022pwc}, including effects like hadronization.

The emerging picture is very consistent with previous investigations \cite{Maas:2017wzi,Maas:2020kda,Maas:2018ska,Dudal:2020uwb,Frohlich:1980gj,Frohlich:1981yi}. The actual physical degrees of freedom are bound states, whose properties are well described using the FMS mechanism. When probed, they behave in a very DIS-like fashion. At low energies the behavior is the one as expected from probing the bound state at a whole, where a typical bound state size of order a few $(\sim\SI{10}{\GeV})^{-1}$ seems to be characteristic, which is much more compact than a typical hadron. At high energies the behavior is given by the FMS-dominating constituent, and is essentially identical to the one in ordinary perturbation theory. This therefore both reproduces the success of perturbation theory and validates its use in this kinematical region. Only towards the TeV range first indications appear that again deviations due to the FMS-subdominant constituents could play a role \cite{Maas:2020kda,Reiner:2021bol,Fernbach:2020tpa,Dudal:2020uwb}.

While it would be very interesting to actually measure the size parameter experimentally, this would require to investigate one of the bound states well separated from its production process, such that it can be considered quasi asymptotically. That appears at least very challenging, given that in the full SM all of these bound states decay. Only the scalar particle inherits a narrow width from the Higgs \cite{Maas:2020kda}, and may therefore live long enough to isolate it well enough. Whether this is experimentally feasible remains to be investigated. A possible avenue has been outlined in \cref{sec:exp}.

Conversely, this implies that corrections to VBS at higher energies may be relatively small, and thus not impede the search for new physics \cite{Cacciapaglia:2020kgq}. But this requires further investigations as well, which are likely only possible within the FMS mechanism as going beyond the inelastic threshold does not yet appear feasible on the lattice given the enormous statistical noise.

There is also a broader implication of the present results. The same reasoning of gauge-invariance also applies to the fermions in the standard model \cite{Frohlich:1980gj,Frohlich:1981yi,Egger:2017tkd} and, by extension, to hadrons \cite{Egger:2017tkd,Fernbach:2020tpa}. For static properties first support in the fermion sector has been gathered from exploratory lattice results as well \cite{Afferrante:2020fhd}. It therefore stands to reason that similar deviations could also arise for leptons\footnote{Note that here an effective custodial/weak radius is searched for, and not the exceedingly small electromagnetic radius \cite{Brodsky:1980zm}.} or hadrons, which do form much better asymptotic states. To estimate the effects will require substantially more involved calculations, and especially on the lattice can only be done by proxy theories so far \cite{Afferrante:2020fhd}. Still, they are a logical and necessary goal, to exclude unaccounted-for standard model background in new physics searches.

\begin{acknowledgments}
We are grateful to C.\ B.\ Lang, S.\ Pl\"atzer, and R.\ Sch\"ofbeck for helpful discussions and to E.\ Dobson and S.\ Pl\"atzer for a critical reading of the manuscript. B.\ R.\ has been supported by the Austrian Science Fund FWF, grant P32760. The computational results presented have been obtained using the HPC center at the University of Graz. We are grateful to its team for its exceedingly smooth operation.
\end{acknowledgments}

\appendix

\section{\label{ap:spectrum}Spectrum extraction}

\begin{figure*}[t!]
    \centering
    \subfloat[$L=8$]{
        \includegraphics[width=0.971\columnwidth]{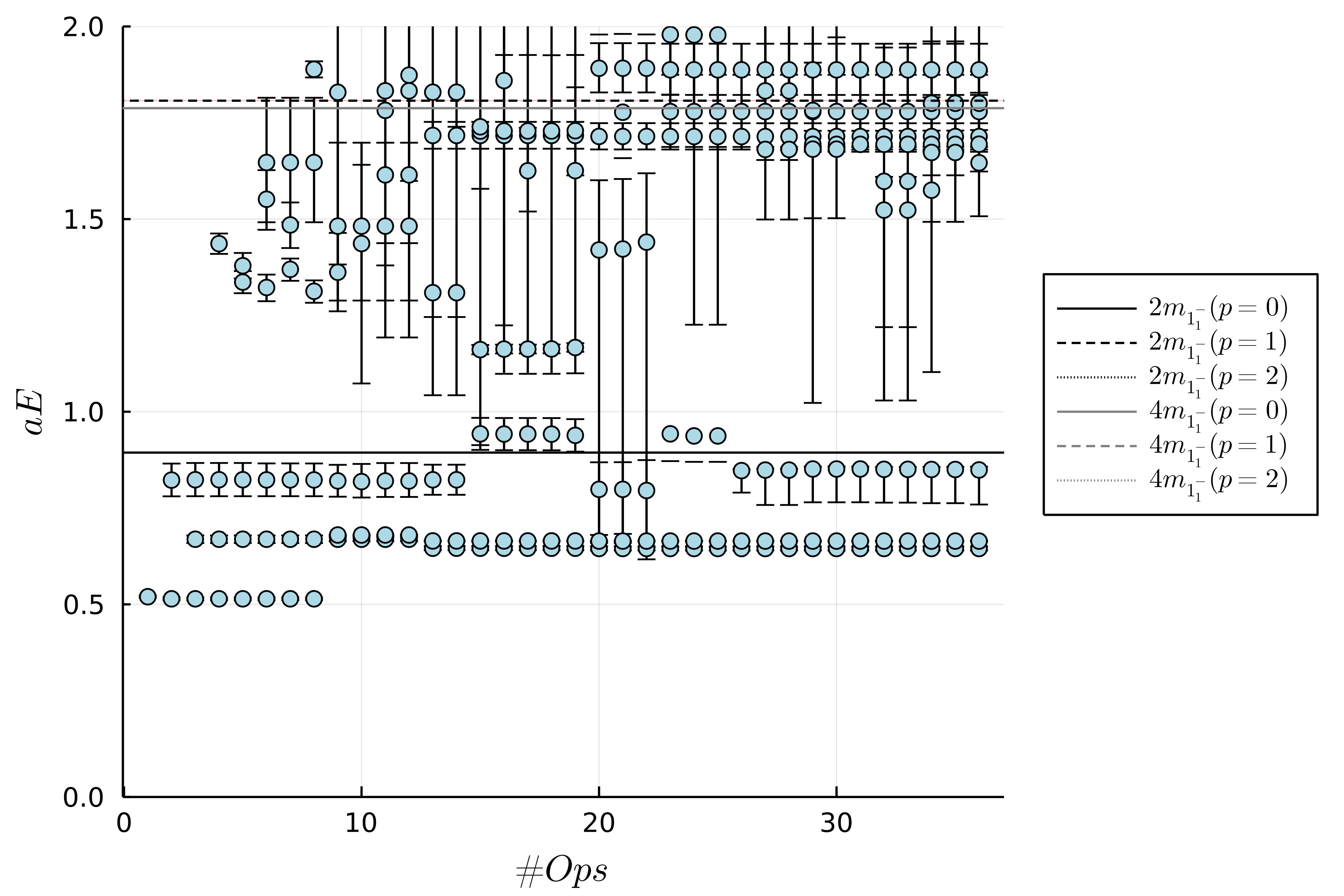}
        \label{fig:ops_l4}
    }
    \hfill
    \subfloat[$L=16$]{
        \includegraphics[width=0.971\columnwidth]{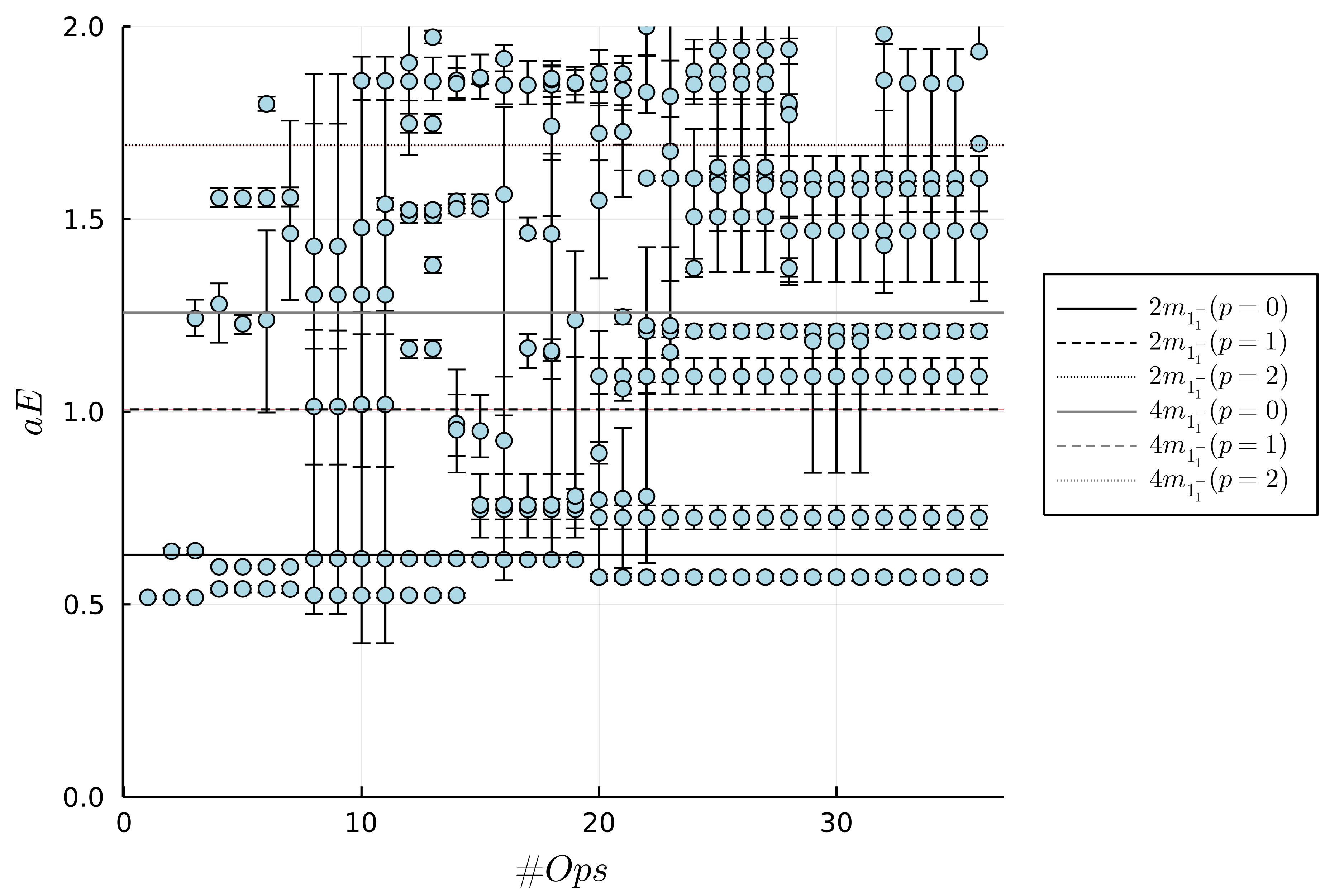}
        \label{fig:ops_l8}
    }
    \\
    \subfloat[$L=24$]{
        \includegraphics[width=0.971\columnwidth]{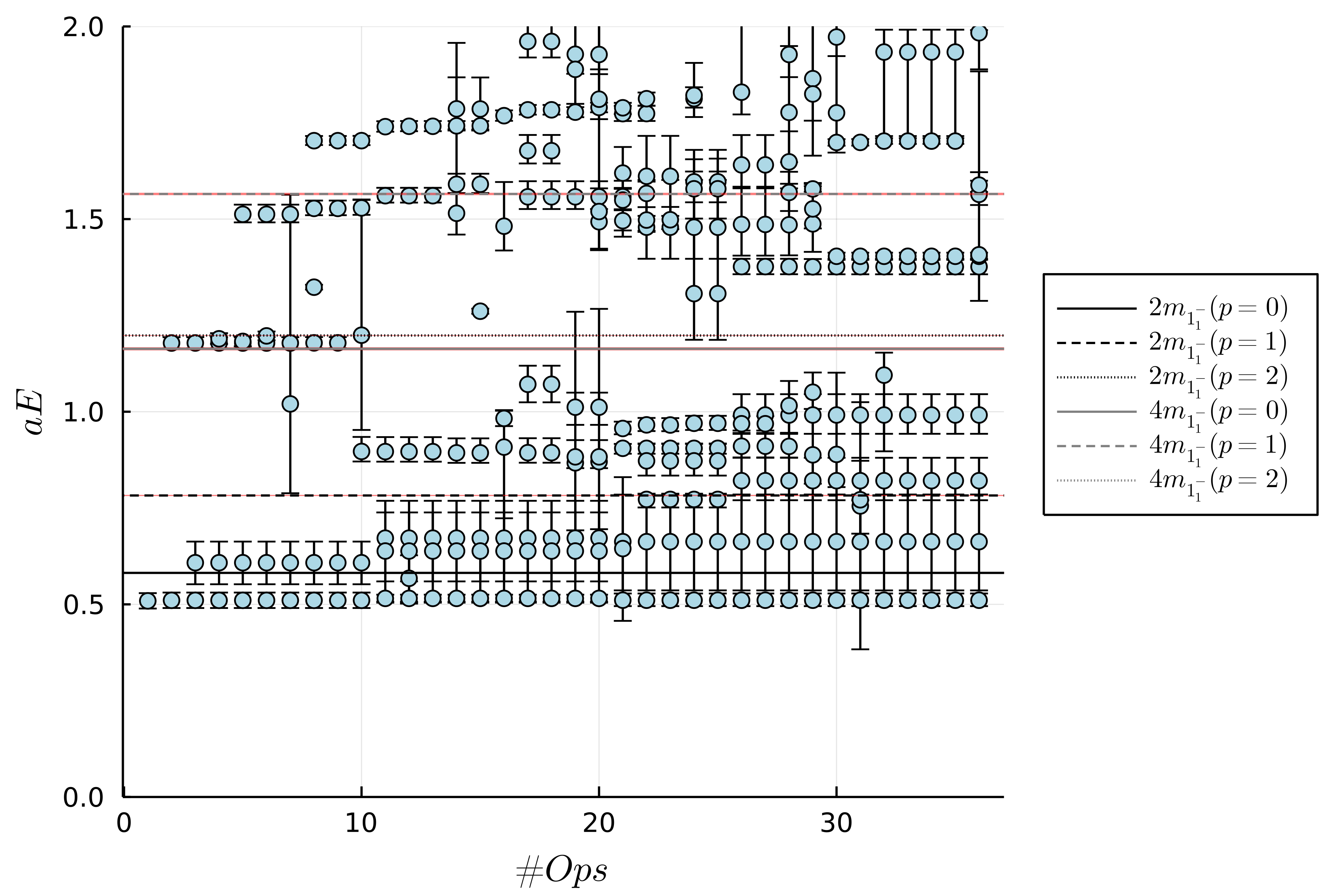}
        \label{fig:ops_l12}
    }
    \hfill
    \subfloat[$L=32$]{
        \includegraphics[width=0.971\columnwidth]{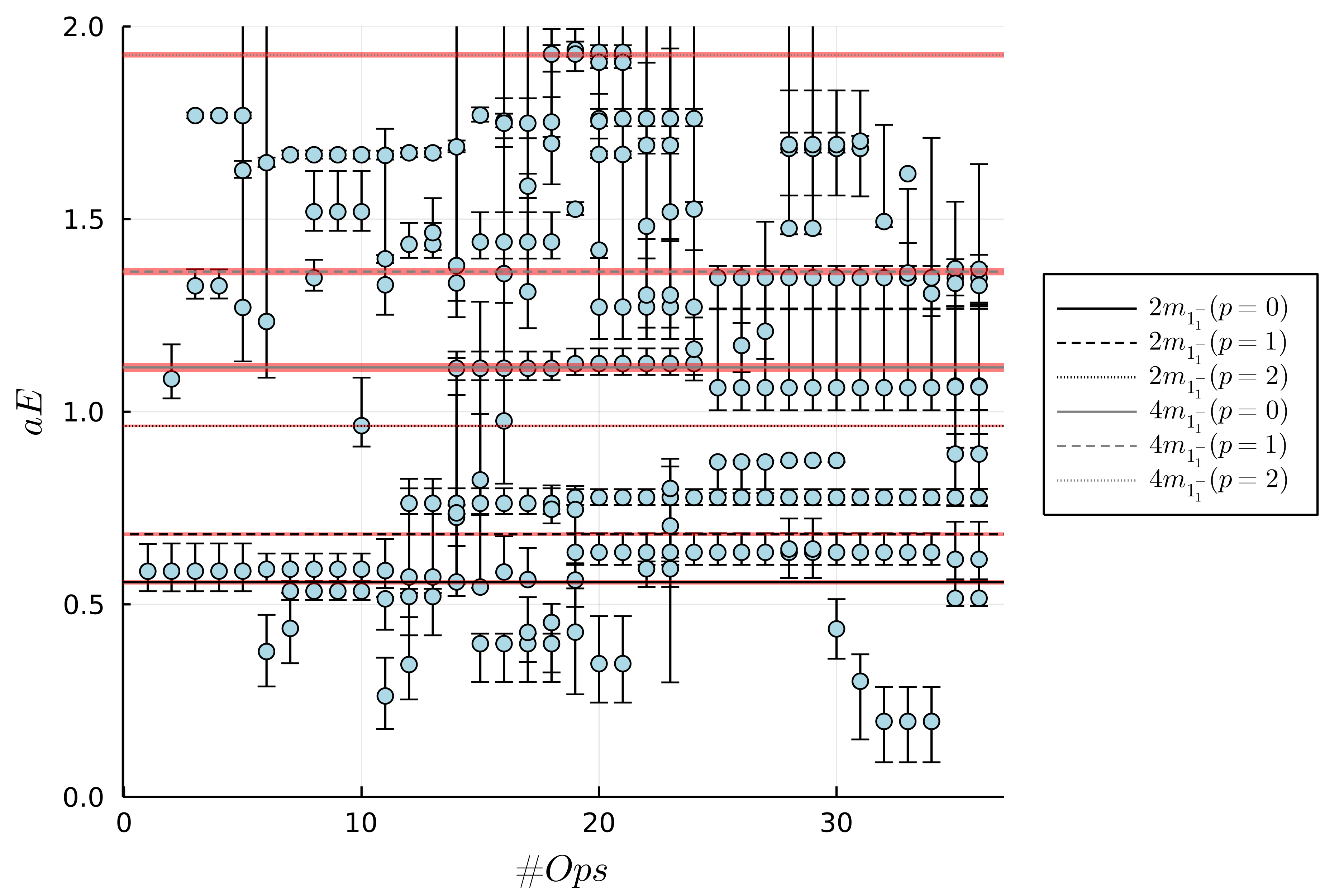}
        \label{fig:ops_l16}
    }
    \caption{Energy levels for set 1 as a function of the operator basis size.}
    \label{fig:volume_dep}
\end{figure*}

\subsection{Energy levels}

In the following, the method to obtain the energy spectrum from the correlator is given in more detail \cite{Jenny:2021}.
The correlator on a lattice with periodic boundary conditions takes the general form
\begin{equation}\label{eqn:corr_on_lat}
    C(t) = \sum_k \, A_k \cosh \left[ E_k \left(t-\frac{L_t}{2}\right)\right].
\end{equation}
Furthermore, it is known that the stochastic noise is independent of $t$. This leads to an exponential increase in the relative uncertainty towards the value at $L_t/2$. In this method, we include only the correlator points until their relative error exceeds \SI{100}{\percent}. This point in time is further referred  to as $t'$ with $t' \leq L_t/2$.

It is now assumed that for the considered correlators the correlator
can in some region be described by a single $\cosh$ term with the ground-state energy $E_0$ and a common value for $A_0$ which
corresponds to the sought-for estimate of $C(L_t/2)$,
\begin{equation}\label{eqn:purecosh}
     C(t) \bigg \vert_{t \in \textrm{Plateau}} = A_0 \cosh \left[ E_0 \left(t-\frac{L_t}{2}\right)\right].
\end{equation}
This region is further referred to as the plateau region.

We now define
\begin{align}\label{eqn:func_Ak_t}
    f\qty(A_0(t),t) &= \left[ \frac{C(t)}{A_0(t)} + \sqrt{ \left( \frac{C(t)}{A_0(t)} \right)^2 - 1 }\right]^{\frac{1}{t-\frac{L_t}{2}}} = \nonumber\\
    &= e^{-E_{eff}(t)}
\end{align}
interrelating the parameters $A_0(t)$, $C(t)$ and $E_{eff}(t)$ to each other, whereby $E_{eff}(t)$ serves a similar purpose as the effective mass introduced in \cref{eqn:effective_energy}. If indeed the correlator would be described by \cref{eqn:purecosh}, $f$ would be a constant.

By demanding $f(A_0(t),t) \stackrel{!}{=} f(A_0(t),t+1)$ a common value of $A_0(t+1/2)$ between two correlator points can be calculated.
These, similar to the function itself, follow a constant behaviour if they are part of the plateau. By taking the mean of the $A_0(t)$s in the would-be constant region, a common value for the whole plateau can be obtained.
After initial tests with a variable plateau region, we could establish that the considered correlators allow a fixed plateau region with $t \in [2,t'-3]$.

The value $\bar{A_0}= \textrm{Mean} \left[ A_0(t \in \textrm{Plateau}) \right]$ is then inserted into \cref{eqn:func_Ak_t}
which again leads to a constant region for $f( \bar{A_0},t)$. By doing a constant fit the final value of the mass
$\bar{E}_{\textrm{Plat.}}$ can be calculated (with $t \in [2,t'-2]$) as
\begin{equation}
  \bar{f} = \textrm{Mean}[f(\bar{A_0},t \in \textrm{Plat.})] \, \rightarrow \,
  \bar{E}_{\textrm{Plat.}} = - \log[\bar{f}].
\end{equation}
Tests with mock-up data and different levels of noise show that this approach provides very good estimates of the actual mass, substantially superior to fits using single or double $\cosh$, and the correlators indeed form plateaus entering the value of $\bar{A_0}$.

\subsection{Selection of operators}

To obtain the full energy spectrum for a given set as a function of volume we performed a variational analysis \cite{Gattringer:2010zz} for each individual lattice size. As already mentioned in the main text we used a variable number of operators in our basis for each set and lattice size respectively.

To have an systematic approach of choosing the operator basis we defined a signal-to-noise ratio $\text{SNR}_i$ for each operator contributing to the correlation matrix by
\begin{equation}
    \text{SNR}_i=\sum_{t=0}^{L_t/2} \frac{\Delta C_{ii}(t)}{C_{ii}(t)}
\end{equation}
with $C_{ij}(t)$ and $\Delta C_{ij}(t)$ as defined in \cref{eqn:correlation_matrix} and \cref{eqn:error_correlation_matrix} respectively.

Starting out with the operator contributing the smallest $\text{SNR}_i$, a first estimate of the ground state energy, using the method from the previous section, has been obtained. Then the operator with the next-smallest $\text{SNR}_i$ has been added to the basis and a variational analysis has been performed to disentangle the states and thus improve the plateau regions. Again the energy levels have been obtained for all eigenvalues using the predictor method. This procedure has then been iterated until all operators were added to the basis. In principle this procedure could also be stopped once a certain total $\text{SNR}$ has been reached, since at some point the variational analysis becomes too noisy.

In \cref{fig:volume_dep} we show the so-obtained energy spectrum for some specific lattice sizes of set 1 as a function of the operator basis size. These figures also contain lines that represent the expected non-interacting states in this region. It can readily be seen from these plots, that always the first operator already predicts quite sufficiently the ground state in this channel. However, we also note, that for the majority of the energy levels the upper and the lower ends of the error bars do exactly coincide with energy levels that are missing for the specific operator basis. This is indeed to be expected since this is basically a remnant of bad disentanglement of close-lying states.

To finally obtain the full spectra as a function of the volume as shown in \cref{fig:spectra} one needs to pick an operator basis for each setup. The number of employed operators as a function of volume is shown in table \cref{tab:nops}. This choice is based on the evolution with the number of operators of the energy levels as shown in figure \ref{fig:volume_dep}. We note that because of levels approaching each others differently on different volume and the general increase of noise with increasing volume, the number of operators for different volumes needs to be done independently.

\subsection{Volume dependence}

For the application of the L\"uscher analysis and also for the use of the perturbative transition amplitudes it is necessary to obtain the infinite volume masses $m_W$ and, if a bound state below threshold is found, $m_H$.
Therefore the resulting data was then fitted by the ansatz \cite{Luscher:1985dn}
\begin{equation}\label{eqn:mvol}
    m_N=m_\infty+\frac{c_0}{L}e^{-c_1L}
\end{equation}
and the errors were obtained again from variation of the input data within $\pm1\sigma$. The resulting parameters can be found in \cref{tab:fits}.

\section{\label{ap:parameters}Lattice parameter sets and fit tables}

The total number of configurations employed in our analysis is given in \cref{tab:statistics}. For reproduce-ability we also give the number of used operators in the variational analysis for each parameter set in \cref{tab:nops}. In \cref{tab:fits} we present all the fit results of the infinite volume masses $m_W$ using \cref{eqn:mvol}, the bound state mass $m_H$ and the coefficients from \cref{eqn:deltaf} in sets 1 and 2 as well as the resonance mass $m_H$ in set 4 using the perturbative prediction of the phase shift. Finally, the energy levels and the corresponding uncertainties in the elastic region, obtained as described in \cref{ap:spectrum}, are given in \crefrange{tab:data_set1}{tab:data_set4} respectively for each parameter set.

\begin{table}[h!]
    \centering
    \caption{Total number of uncorrelated configurations per data set and lattice size.}
    \label{tab:statistics}
    \begin{tabular}{r|r|r|r|r}
        L  & Set 1  & Set 2  & Set 3  & Set 4   \\ \hline\hline
        8  & 664 724 & 555 881 & 828 985 & 1 412 021 \\
        12 & 463 852 & 409 854 & 481 642 &   481 642 \\
        16 & 656 962 & 511 968 & 781 562 &   751 500 \\
        20 & 452 231 & 434 070 & 496 460 &   512 148 \\
        24 & 629 442 & 579 293 & 682 715 &   848 591 \\
        28 & 446 666 & 426 794 & 320 115 &   381 307 \\
        32 & 105 371 & 107 035 &  89 428 &    82 626
    \end{tabular}
\end{table}
\begin{table}[h!]
    \vspace{-2em}
    \centering
    \caption{Chosen number of operators for the variational analysis to obtain the energy levels in \crefrange{tab:data_set1}{tab:data_set4}.}
    \label{tab:nops}
    \begin{tabular}{r|r|r|r|r}
         L  & Set 1 & Set 2 & Set 3 & Set 4\\\hline\hline
         8  &   8 & 11 & 11 &  9 \\
         12 &  13 & 13 & 10 & 10 \\
         16 &  13 &  4 & 11 & 16 \\
         20 &   9 & 11 &  9 & 14 \\
         24 &  12 & 11 & 20 & 11 \\
         28 &  10 & 11 & 23 & 11 \\
         32 &  13 & 10 & 13 &  8
    \end{tabular}
\end{table}
\begin{table}[h!]
    \vspace{-2em}
    \centering
    \caption{Fit results for the parameters of the given equations. Values are given in terms of inverse lattice spacing except for $c_1$, which is dimensionless.}
    \label{tab:fits}
    \bgroup
    \renewcommand{\arraystretch}{1.5}
    \begin{tabular}{c|c|c|c|c|c}
        Fit par. & eqn. & Set 1  & Set 2  & Set 3  & Set 4   \\ \hline\hline
        $m_W$ & \ref{eqn:mvol} & $0.280(3)$ & $0.277(3)$ & $0.278(2)$ & $0.330(5)$ \\
        $c_0$ & \ref{eqn:mvol} & $3.4(2)$ & $3.3(2)$ & $2.4(3)$ & $1.5(7)$ \\
        $c_1$ & \ref{eqn:mvol} & $0.118(6)$ & $0.126(6)$ & $0.18(2)$ & $0.17(6)$ \\\hline
        $m_H$ & \ref{eqn:mvol}/\ref{eqn:pert_tan} & $0.52_{-0.07}^{+0.02}$ & $0.52_{-0.04}^{+0.02}$ & -- & $1.134(11)$\\
        $c_0$ & \ref{eqn:mvol} & $0.6^{+0.2}_{-0.5}$ & $0.30^{+0.14}_{-0.05}$ & -- & -- \\
        $c_1$ & \ref{eqn:mvol} & $0(3)$ & $0(3)$ & -- & -- \\\hline
        $a_0^{-1}$ & \ref{eqn:deltaf} & $0.136_{-0.021}^{+0.035}$ & $0.041_{-0.028}^{+0.035}$ & -- & -- \\
        $a_1^{-1}$ & \ref{eqn:deltaf} & $0.143_{-0.004}^{+0.007}$ & $0.148_{-0.011}^{+0.018}$ & -- & --
    \end{tabular}
    \egroup
\end{table}
\begin{table}
    \centering
    \caption{Data 2.7984-0.2984-1.317}
    \label{tab:data_set1}
    \begin{tabular}{c|r|r|r}
        L & $E\ \qty[\si{\GeV}]$ & $E^{-}\ \qty[\si{\GeV}]$ & $E^{+}\ \qty[\si{\GeV}]$ \\ \hline\hline
        8  & 147 & 1  & 1  \\
        8  & 192 & 3  & 3  \\
        8  & 236 & 13 & 13 \\ \hline
        12 & 156 & 2  & 2  \\
        12 & 183 & 3  & 3  \\
        12 & 268 & 18 & 17 \\ \hline
        16 & 150 & 2  & 2  \\
        16 & 177 & 3  & 3  \\ \hline
        20 & 149 & 4  & 4  \\
        20 & 173 & 7  & 7  \\ \hline
        24 & 148 & 4  & 3  \\
        24 & 163 & 19 & 18 \\
        24 & 183 & 38 & 29 \\
        24 & 193 & 33 & 28 \\
        24 & 257 & 8  & 11 \\ \hline
        28 & 147 & 15 & 15 \\
        28 & 171 & 18 & 16 \\
        28 & 295 & 19 & 58 \\ \hline
        32 & 149 & 29 & 6  \\
        32 & 164 & 12 & 74 \\
        32 & 219 & 8  & 12
    \end{tabular}
\end{table}
\begin{table}
    \centering
    \caption{Data 2.8859-0.2981-1.334}
    \label{tab:data_set2}
    \begin{tabular}{c|r|r|r}
        L & $E\ \qty[\si{\GeV}]$ & $E^{-}\ \qty[\si{\GeV}]$ & $E^{+}\ \qty[\si{\GeV}]$ \\ \hline\hline
        8  & 146 & 1   & 1   \\
        8  & 202 & 2   & 2   \\
        8  & 250 & 4   & 4   \\ \hline
        12 & 155 & 2   & 2   \\
        12 & 193 & 3   & 3   \\
        12 & 254 & 7   & 7   \\
        12 & 671 & 451 & 110 \\ \hline
        16 & 150 & 3   & 3   \\
        16 & 188 & 5   & 5   \\
        16 & 240 & 4   & 4   \\ \hline
        20 & 149 & 4   & 4   \\
        20 & 185 & 8   & 8   \\
        20 & 254 & 32  & 35  \\ \hline
        24 & 149 & 6   & 6   \\
        24 & 184 & 14  & 14  \\
        24 & 224 & 7   & 8   \\
        24 & 298 & 33  & 40  \\ \hline
        28 & 147 & 16  & 13  \\
        28 & 181 & 29  & 31  \\
        28 & 207 & 6   & 6   \\
        28 & 288 & 26  & 60  \\ \hline
        32 & 150 & 7   & 15  \\
        32 & 165 & 26  & 37  \\
        32 & 198 & 13  & 14  \\
        32 & 209 & 9   & 9
    \end{tabular}
\end{table}
\begin{table}
    \centering
    \caption{Data 4.0-0.285-0.97}
    \label{tab:data_set3}
    \begin{tabular}{c|r|r|r}
        L & $E\ \qty[\si{\GeV}]$ & $E^{-}\ \qty[\si{\GeV}]$ & $E^{+}\ \qty[\si{\GeV}]$ \\ \hline\hline
        8  & 144 & 1   & 1   \\
        8  & 278 & 3   & 3   \\
        8  & 300 & 3   & 3   \\ \hline
        12 & 142 & 1   & 1   \\
        12 & 283 & 9   & 9   \\
        12 & 305 & 4   & 4   \\
        12 & 346 & 72  & 95  \\ \hline
        16 & 145 & 2   & 2   \\
        16 & 255 & 9   & 9   \\
        16 & 308 & 17  & 17  \\ \hline
        20 & 148 & 4   & 5   \\
        20 & 224 & 16  & 15  \\
        20 & 272 & 27  & 29  \\
        20 & 287 & 31  & 7   \\ \hline
        24 & 151 & 3   & 5   \\
        24 & 215 & 8   & 9   \\
        24 & 267 & 6   & 6   \\
        24 & 290 & 6   & 61  \\ \hline
        28 & 129 & 59  & 57  \\
        28 & 146 & 15  & 6   \\
        28 & 173 & 29  & 45  \\
        28 & 252 & 2   & 1   \\
        28 & 303 & 20  & 11  \\
        28 & 343 & 29  & 50  \\
        28 & 991 & 847 & 847 \\ \hline
        32 & 160 & 13  & 8   \\
        32 & 193 & 41  & 24  \\
        32 & 195 & 16  & 59
    \end{tabular}
\end{table}
\begin{table}
    \centering
    \caption{Data 4.0-0.3-1.0}
    \label{tab:data_set4}
    \begin{tabular}{c|r|r|r}
        L & $E\ \qty[\si{\GeV}]$ & $E^{-}\ \qty[\si{\GeV}]$ & $E^{+}\ \qty[\si{\GeV}]$ \\ \hline\hline
        8  & 141 & 1   & 1   \\
        8  & 276 & 3   & 3   \\ \hline
        12 & 143 & 1   & 1   \\
        12 & 278 & 5   & 5   \\
        12 & 289 & 22  & 13  \\
        12 & 375 & 116 & 133 \\ \hline
        16 & 148 & 2   & 3   \\
        16 & 232 & 8   & 8   \\
        16 & 265 & 67  & 24  \\
        16 & 299 & 45  & 57  \\ \hline
        20 & 155 & 6   & 12  \\
        20 & 211 & 20  & 15  \\
        20 & 278 & 6   & 6   \\
        20 & 325 & 48  & 30  \\ \hline
        24 & 153 & 12  & 17  \\
        24 & 202 & 7   & 8   \\
        24 & 220 & 44  & 52  \\ \hline
        28 & 156 & 18  & 20  \\
        28 & 190 & 14  & 16  \\
        28 & 214 & 9   & 16  \\
        28 & 319 & 23  & 20  \\ \hline
        32 & 162 & 8   & 6   \\
        32 & 184 & 9   & 16  \\
        32 & 221 & 1   & 1
    \end{tabular}
\end{table}

\section{\label{ap:zeta}Generalized Zeta-function $\mathcal{Z}_{Jm}^{\va{d}}\qty(r,q^2)$}

Here we want to present the analytical continuation and numerical evaluation of the zeta function $\mathcal{Z}_{Jm}^{\va{d}}\qty(r,q^2)$ defined in \cref{eqn:zeta} we employed. This has already been discussed by L\"uscher in \cite{Luscher:1990ux,Luscher:1985dn} and on several other occasions \cite{Feng:2010jyb,Gottlieb:1995dk,Luu:2011ep,Gockeler:2012yj,Morningstar:2017spu} for different systems, i.e. moving reference frames. Here we follow \cite{CP-PACS:2004dtj}, where a faster converging version is presented. However, it turned out that there are some typographical errors and discrepancies between the published and the preprint version of \cite{CP-PACS:2004dtj}. Therefore, for completeness and consistency of our own results, we present a full derivation of the numerically stable, analytically continued zeta function in the rest frame ($\va{d}=\va{0}$) for the case of spinless particles ($J=m=0$) employed by us, following the steps of \cite{CP-PACS:2004dtj}. For comparison we will also state the slower converging formula derived in \cite{Feng:2010jyb}.

The definition of the zeta function $\mathcal{Z}_{00}^{\va{d}}\qty(r,q^2)$ is
\begin{gather}\label{eqn:ap_zeta}
    \mathcal{Z}_{Jm}^{\va{d}}\qty(r,q^2) = \sum_{\va{x}\in P_{\va{d}}}
    \frac{\qty|\va{x}|^J Y_{Jm}\qty(\va{x})}{\qty(\va{x}^2-q^2)^{r}} \\\nonumber
    P_{\va{d}} = \Set{\va{x} \in \mathbb{R}^3 | \va{x} = \va{y} + \frac{\va{d}}{2}, \va{y} \in \mathbb{Z}^3 } \\
    \intertext{and takes on finite values for $\Re{r} > 3/2$. For the spinless case \cref{eqn:ap_zeta} simplifies to}
   \mathcal{Z}_{00}^{\va{d}}\qty(r,q^2) = \frac{1}{\sqrt{4\pi}} \sum_{\va{x}\in P_{\va{d}}} \frac{1}{\qty(\va{x}^2-q^2)^{r}}.
\end{gather}
 To calculate scattering phase shifts as shown in \cref{eqn:phase shift} we need $r=1$ and thus an analytic continuation. First we split the sum into two parts
\begin{equation}\label{eqn:ap_split}
    \sum_{\va{x}} \frac{1}{\qty(\va{x}^2-q^2)^{r}} = \sum_{\va{x}^2 < q^2} \frac{1}{\qty(\va{x}^2-q^2)^{r}} + \sum_{\va{x}^2 > q^2} \frac{1}{\qty(\va{x}^2-q^2)^{r}}
\end{equation}
with $\va{x}\in P_{\va{d}}$. For the second sum, the denominator is always larger zero and thus it is possible to rewrite it in the following way
\begin{widetext}
\begin{subequations}
    \begin{align}\nonumber\label{eqn:gamma_pos_a}
    \sum_{\va{x}^2 > q^2} \frac{1}{\qty(\va{x}^2-q^2)^{r}} &= \sum_{\va{x}^2 > q^2} \frac{1}{\Gamma\qty(r)} \int_{0}^{\infty} \frac{\dd{t}}{\qty(\va{x}^2-q^2)} \qty(\frac{t}{\va{x}^2-q^2})^{r-1} e^{-t}
    \qquad \text{substitute } u = \frac{t}{\va{x}^2-q^2}\\
    &= \frac{1}{\Gamma\qty(r)}  \sum_{\va{x}^2 > q^2} \qty[\int_{0}^{1} \dd{u} u^{r-1} e^{-u\qty(\va{x}^2-q^2)} + \int_{1}^{\infty} \dd{u} u^{r-1} e^{-u\qty(\va{x}^2-q^2)}] \\
    \intertext{Adding and again subtracting the $\va{x}^2<q^2$ part of this sum for the first integral and reverting the substitution for some parts yields the following equation}\nonumber\label{eqn:gamma_pos_b}
    &= \frac{1}{\Gamma\qty(r)} \int_{0}^{1} \dd{u} u^{r-1} e^{uq^2}\sum_{\va{x}} e^{-u\va{x}^2} - \frac{1}{\Gamma\qty(r)}\sum_{\va{x}^2 < q^2}\frac{\Gamma\qty(r)-\Gamma\qty(r,\va{x}^2-q^2)}{\qty(\va{x}^2-q^2)^{r}}\\
    &+ \frac{1}{\Gamma\qty(r)}\sum_{\va{x}^2 > q^2} \int_{1}^{\infty} \dd{u} u^{r-1} e^{-u\qty(\va{x}^2-q^2)} \\
    \intertext{with $\Gamma\qty(r,\va{x}^2-q^2)$ the upper incomplete gamma function. Although the second argument of this function is a negative value, which in general is not well-defined, this is still valid in our case due to the denominator. Note also that the first sum now covers the full range of $\va{x}\in P_{\va{d}}$ again. Finally the last integral can be solved by again reverting the substitution. Collecting all terms together we end up with the following result}\label{eqn:gamma_pos_c}
    &= \frac{1}{\Gamma\qty(r)} \int_{0}^{1} \dd{u} u^{r-1} e^{uq^2}\sum_{\va{x} } e^{-u\va{x}^2} - \sum_{\va{x}^2 < q^2} \frac{1}{\qty(\va{x}^2-q^2)^{r}} + \sum_{\va{x}}\sum_{k=1}^{r} \frac{e^{-\qty(\va{x}^2-q^2)}}{\qty(r-k)!\qty(\va{x}^2-q^2)^{k}}.
\end{align}
\end{subequations}
\end{widetext}
where the second term in \cref{eqn:gamma_pos_c} exactly cancels the first term in \cref{eqn:ap_split}. Our final form for this equation turns out to be a combination of the versions stated in the published and preprint version of \cite{CP-PACS:2004dtj}.

For further evaluation we can use Poisson's summation formula
\begin{equation}\label{eqn:ap_poisson}
    \sum_{\va{z}\in\mathbb{Z}^3}f\qty(\va{z}) = \sum_{\va{z}\in\mathbb{Z}^3}\int \dd[3]{r}f\qty(\va{r})e^{i2\pi\va{z}\va{r}}
\end{equation}
to simplify the first integral in \cref{eqn:gamma_pos_c}. By integrating over $\va{r}$ and explicitly inserting $\va{x}$ from \cref{eqn:ap_zeta} we arrive at
\begin{equation}\label{eqn:term1}
    \sum_{{\va{x}\in P_{\va{d}}}} e^{-u\va{x}^2} = \qty(\frac{\pi}{u})^{\frac{3}{2}}\sum_{\va{y}\in\mathbb{Z}^3} \qty(-1)^{\va{y}\va{d}}e^{-\frac{\pi^2\va{y}^2}{u}}
\end{equation}
for the sum inside the integral.
The divergence at $r=1$ is due to the $\va{y}=\va{0}$ term in the sum. Splitting the sum in a divergent ($\va{y}=\va{0}$) and a finite part ($\va{y}\neq\va{0}$) finally allows us to separate the divergence and to analytically continue the function
\begin{align} \label{eqn:ap_term_diverging}\nonumber
    \int_{0}^{1} \dd{u} u^{r-1} e^{uq^2}\qty(\frac{\pi}{u})^{\frac{3}{2}} &= \pi^{\frac{3}{2}}\sum_{k=0}^{\infty} \frac{\qty(q^2)^k}{k!} \int_{0}^{1} \dd{u} u^{r-\frac{3}{2}+k-1} \\
    &= \pi^{\frac{3}{2}}\sum_{k=0}^{\infty} \frac{\qty(q^2)^k}{k!} \frac{1}{r+k-3/2}
\end{align}
which only works for $\Re\qty{r}>3/2$. However, the right hand side takes a finite value for $r=1$ and thus can be used for continuation. The final expression for the zeta function for $r=1$ is
\begin{align}
    \sqrt{4\pi}\mathcal{Z}_{00}^{\va{d}}\qty(1,q^2) &= \sum_{{\va{x}\in P_{\va{d}}}} \frac{e^{-\qty(\va{x}^2-q^2)}}{\qty(\va{x}^2-q^2)} + \pi^{\frac{3}{2}}\sum_{k=0}^{\infty} \frac{\qty(q^2)^k}{k!} \frac{1}{k-1/2} + \nonumber\\
    &+ \int_{0}^{1} \dd{u}  e^{uq^2}\qty(\frac{\pi}{u})^{\frac{3}{2}}\sum_{\va{y}\in\mathbb{Z}^3}{}^{'} \qty(-1)^{\va{y}\va{d}}e^{-\frac{\pi^2\va{y}^2}{u}} \label{eqn:ap_zeta_full}
\end{align}
where the summation $\sum_{\va{y}\in\mathbb{Z}^3}{}^{'}$ intends that $\va{y}=\va{0}$ has been left out.

Inserting also $\va{d}=\va{0}$ into the formula yields finally a numerically stable and fast converging representation of the generalized zeta function as needed in \cref{eqn:phase shift}.
\begin{align}
    \sqrt{4\pi}\mathcal{Z}_{00}^{\va{0}}\qty(1,q^2) &= \sum_{\va{y}\in\mathbb{Z}^3} \frac{e^{-\qty(\va{y}^2-q^2)}}{\qty(\va{y}^2-q^2)} + \pi^{\frac{3}{2}}\sum_{k=0}^{\infty} \frac{\qty(q^2)^k}{k!} \frac{1}{k-1/2} + \nonumber\\
    &+ \int_{0}^{1} \dd{u}  e^{uq^2}\qty(\frac{\pi}{u})^{\frac{3}{2}}\sum_{\va{y}\in\mathbb{Z}^3}{}^{'} e^{-\frac{\pi^2\va{y}^2}{u}}. \label{eqn:ap_zeta_final}
\end{align}
Note, that here the first sum \textbf{does include} $\va{y}=\va{0}$ not as stated in the published version of \cite{CP-PACS:2004dtj}. That \cref{eqn:ap_zeta_final} is indeed the correct form can be seen by comparison with \cite{Gottlieb:1995dk,Feng:2010jyb}.

The implementation used in this work follows \cref{eqn:ap_zeta_final} and has been verified by comparison with values obtained as described in \cite{Feng:2010jyb}. In this paper a different expression for \cref{eqn:ap_term_diverging} is given by
\begin{equation}\label{eqn:ap_term_diverging_feng}
    \int_{0}^{1} \dd{u} u^{r-1} e^{uq^2}\qty(\frac{\pi}{u})^{\frac{3}{2}} = -2\pi^{\frac{3}{2}} + \int_{0}^{1} \dd{u} \qty(e^{uq^2} - 1)\qty(\frac{\pi}{u})^{\frac{3}{2}}.
\end{equation}
However, both yielded numerically the same results but the integral version in \cref{eqn:ap_term_diverging_feng} converged much slower. To verify the calculations in this work, the zeros of the generalized zeta function have been calculated and compared with those stated in \cite{Gockeler:1994rx}.

\section{\label{ap:pt}Relativistic kinematics, perturbative phase shifts and transition amplitudes}

The kinematic assignments in this work follow closely the ones in \cite{Denner:1996ug,Denner:1997kq}. However, keeping in mind that the prescriptions have to be compared to lattice-data it is more instructive to use the following conventions for the dispersion relation and the Mandelstam variables.
\begin{align}
    p^2 &= \qty(\frac{E}{2})^2-m_W^2\\
    s&=E^2\\
    t &= -4p^2\sin\qty(\frac{\theta}{2})^2\\
    u &= -4p^2\cos\qty(\frac{\theta}{2})^2
\end{align}
with $\theta$ being the scattering angle between the incoming and the outgoing particles. This conventions only differ to those in \cite{Denner:1996ug,Denner:1997kq} by a replacement of $E$ by $E/2$.

Using this conventions we obtained the polarised transition amplitudes given in \cref{tab:ap_amplitudes}.

The full amplitude from \cref{eqn:matrixelement} in the Born approximation can be obtained from the individual elements in \cref{tab:ap_amplitudes}. It is therefore given by
\begin{widetext}
    \bgroup
    \allowdisplaybreaks
\begin{align}\label{eqn:ap_mborn}
    \mathcal{M}_B = \frac{g_W^2}{36}  &\left\{\frac{\qty(1-2 c^2) \qty(E^2+4 m_W^2)^2}{m_W^4} +\frac{8 \qty(18 E^6-131 E^4 m_W^2+520 E^2m_W^4-816 m_W^6)}{\qty(E^2-4 m_W^2)^2 \qty((1-c) E^2+2(2c-1) m_W^2)} -\frac{4 c \qty(5 E^2+4 m_W^2)^2}{E^4-5 E^2 m_W^2+4 m_W^4} \right.\nonumber\\
    & +\frac{10 \qty(m_H^4 \qty(E^2+4 m_W^2)^2+4 m_H^2 \qty(3 E^4 m_W^2-8 E^2 m_W^4-16 m_W^6)+12 m_W^4 \qty(E^2-4m_W^2)^2)}{m_W^2\qty(E^2-4 m_W^2)^2 \qty(2 m_H^2+(1-c) \qty(E^2-4 m_W^2))}  \nonumber\\
    &+\frac{2\qty( m_H^4 \qty(E^2+4 m_W^2)^2+4 m_H^2 \qty(3 E^4 m_W^2-8 E^2 m_W^4-16 m_W^6)+12 m_W^4\qty(E^2-4 m_W^2)^2)}{m_W^2 \qty(E^2-4 m_W^2)^2 \qty(2 m_H^2 + (1+c) \qty(E^2-4 m_W^2))} \nonumber\\
    &+\frac{1}{m_W^4 \qty(E^2-m_H^2) \qty(E^2-4 m_W^2)^2} \left[5 E^{10}-44E^8 m_W^2-60 E^6 m_W^4+412 E^4 m_W^6-416 E^2 m_W^8-64 m_W^{10}\right. \nonumber\\
    &\left.\left.+6 m_H^4 m_W^2 \qty(E^2+4 m_W^2)^2+m_H^2 \qty(-5 E^8+37 E^6 m_W^2+24 E^4 m_W^4-560 E^2 m_W^6+512 m_W^8) \right] \vphantom{\frac{\qty(1_1)^2}{\qty(1_1)^2}}\right\}
\end{align}
\egroup
where $c=\cos\qty(\theta)$. $\mathcal{M}_B $ can be split into $\mathcal{M}_B^W + \mathcal{M}_B^{H}$, with $\mathcal{M}_B^W$ collecting the terms that are independent of $m_H$ (i.e.\ the first line of \cref{eqn:ap_mborn}) and $\mathcal{M}_B^H$ the rest. For the case when there is no Higgs found below or in the elastic region (as is the case for set 3) the diagrams in \cref{fig:transitionelements} containing Higgs exchanges need to be dropped. This can be achieved by calculating the limit $m_H\to\infty$ of \cref{eqn:ap_mborn}. In this case $\mathcal{M}_B^W$ is unaffected and the remaining term simplifies to
\begin{equation}\label{eqn:born_woh}
    \lim_{m_H\to\infty}\mathcal{M}_B^H = \frac{g_W^2}{36} \frac{2 (c-23) E^6 m_W^2+4 (2 c-3) E^4 m_W^4-32 (c-10) E^2 m_W^6-64 (2 c+11) m_W^8+5 E^8}{m_W^4\qty(E^2-4m_W^2)^2}
\end{equation}

The final equations for the partial wave transition amplitudes $f_0$ in \cref{eqn:partial_amplitudes} and the integrated cross section $\sigma_{\text{Born}}$ in \cref{eqn:xs} are not given here since these are quite involved expressions but can be straightforwardly obtained by evaluating the integrals
\begin{align}
    f_0 = \frac{1}{32\pi}\int_{-1}^1 \mathcal{M}_B \dd{c}  & &
    \sigma_{\text{Born}} = \frac{1}{64\pi^2E^2}\int_{-1}^1 \abs{\mathcal{M}_B}^2 \dd{c}.
\end{align}

\begin{table*}[ht!]
    \centering
    \caption{Polarised transition amplitudes for the tree-level processes in the elastic region of the $0_0^{+}$ channel. We use $c=\cos\qty(\theta)$  as an abbreviation. All entries need to be multiplied by the overall factor $g_W^2=4\pi\alpha_W$.}
    \label{tab:ap_amplitudes}
    \renewcommand{\arraystretch}{2.45}
    \begin{tabular}{c|c}
        Polarization & $\mel{\text{WW}}{T}{\text{WW}}_{pol}$ \\\hline\hline
        ${\{0,0,0,0\}}$ & $\frac{\qty(c^2+6 c-3) E^4}{16 m_W^4}+\frac{8 (1-3 c) E^2}{16 m_W^2}-\frac{\qty(4 p^2-c E^2)^2}{16 m_W^2 \qty(t-m_H^2)}-\frac{\qty(E^2-2 m_W^2)^2}{4 m_W^2 \qty(s-m_H^2)}-\frac{(E^2-u) \qty(2 c m_W^2+t)^2}{4 m_W^4 \qty(t-m_W^2)}+\frac{4 E^2 t (1+3c)}{4 m_W^2 \qty(t-m_W^2)}-\frac{c p^2 \qty(E^2+2 m_W^2)^2}{m_W^4 \qty(s-m_W^2)}$ \\
        ${\{\pm ,\mp ,0,0\}}$ & $\frac{1-c^2}{8} \qty[\frac{E^2}{m_W^2}+\frac{E^2 \qty(E^2+u)}{m_W^2 \qty(t-m_W^2)}-\frac{E^2}{t-m_H^2}-\frac{4 \qty(E^2+4 m_W^2)}{t-m_W^2}]$ \\
        ${\{\pm ,\mp ,\pm ,\mp \}}$ & $\frac{(1+c)^2}{4}  \qty(\frac{t-m_H^2-m_W^2}{t-m_H^2}+\frac{u-E^2}{t-m_W^2})$ \\
        ${\{\pm ,\mp ,\mp ,\pm \}}$ & $\frac{(1-c)^2}{4}  \qty(\frac{t-m_H^2-m_W^2}{t-m_H^2}+\frac{u-E^2}{t-m_W^2})$ \\
        \hline & $\mel{\text{WW}}{T}{\text{ZZ}}_{pol}$ \\\hline\hline
        ${\{0,0,0,0\}}$ & $\frac{2 \qty(c^2-3) E^4+12 E^2 m_W^2}{16 m_W^4}+\frac{4 m_W^2 \qty(E^2-m_W^2)-E^2 m_H^2}{4 m_W^2 \qty(s-m_H^2)}$ \\
        ${\{\pm ,\mp ,0,0\}}$ & $\frac{E^2\qty(1-c^2)}{4 m_W^2}$ \\
        ${\{\pm ,\mp ,\pm ,\mp \}}$ & $\frac{\qty(1+c)^2}{2}$  \\
        ${\{\pm ,\mp ,\mp ,\pm \}}$ & $\frac{\qty(1-c)^2}{2}$ \\
        \hline & $\mel{\text{ZZ}}{T}{\text{ZZ}}_{pol}$ \\\hline\hline
        ${\{0,0,0,0\}}$ & $\frac{\qty(E^2-2 m_W^2)^2}{4 m_W^2\qty(s-m_H^2)}+\frac{\qty(E^2 (1-c)-4 m_W^2)^2}{16 m_W^2\qty(t-m_H^2)}+\frac{\qty(E^2 (1+c)-4 m_W^2)^2}{16 m_W^2\qty(u-m_H^2)}$ \\
        ${\{\pm ,\mp ,0,0\}}$ & $\frac{E^2\qty(1-c^2)}{4}\frac{2p^2+m_H^2}{\qty(t-m_H^2)\qty(u-m_H^2)}$ \\
        ${\{\pm ,\mp ,\pm ,\mp \}}$ & $\frac{m_W^2\qty(1+c)^2}{2}\frac{2p^2+m_H^2}{\qty(t-m_H^2)\qty(u-m_H^2)}$  \\
        ${\{\pm ,\mp ,\mp ,\pm \}}$ & $\frac{m_W^2\qty(1-c)^2}{2}\frac{2p^2+m_H^2}{\qty(t-m_H^2)\qty(u-m_H^2)}$  \\
        \hline
    \end{tabular}
\end{table*}
\end{widetext}

\bibliography{literature}

\end{document}